\documentclass[prd, twocolumn]{revtex4}

\usepackage{epsfig}
\usepackage{graphicx}
\usepackage{amsmath}
\usepackage{amsfonts}
\usepackage{epstopdf}
\usepackage{array}

\renewcommand{\vec}{\boldsymbol}
\newcommand{\rmi}{\mathrm{i}}
\newcommand{\rme}{\mathrm{e}}
\newcommand{\rmd}{\mathrm{d}}

\begin{document}

\title{Vortex formation in a rotating two-component Fermi gas}

\author{Harmen J. Warringa}
\email{warringa@th.physik.uni-frankfurt.de}

\author{Armen Sedrakian}
\email{sedrakian@th.physik.uni-frankfurt.de}
\affiliation{Institut f\"ur Theoretische Physik, Goethe-Universit\"at
  Frankfurt am Main, Max-von-Laue-Stra\ss e~1, 60438 Frankfurt am Main, Germany}

\newcommand{\sumint}{\sum \!\!\!\!\!\!\!\! \int \;}

\date{\today}

\begin{abstract}
  A two-component Fermi gas with attractive $s$-wave interactions
  forms a superfluid at low temperatures. When this gas is confined in
  a rotating trap, fermions can unpair at the edges of the gas and
  vortices can arise beyond certain critical rotation frequencies.  We
  compute these critical rotation frequencies and construct the phase
  diagram in the plane of scattering length and rotation frequency for
  different total number of particles. We work at zero temperature and
  consider a cylindrically symmetric harmonic trapping potential.  The
  calculations are performed in the Hartree-Fock-Bogoliubov
  approximation which implies that our results are quantitatively
  reliable for weak interactions.
\end{abstract}

\maketitle

\section{Introduction}
A characteristic feature of superfluids is the appearance of vortices
when they are rotated. This fact has been used to demonstrate that a
two-component atomic Fermi gas becomes a superfluid at sufficiently low
temperatures \cite{Zwierlein05, Zwierlein06, Zwierlein06b, Schunck07}.
A superfluid state can be created in such a gas by trapping fermionic
atoms in two distinct hyperfine states. The interaction strength
between the two components can be controlled by an external magnetic
field. When the interactions are tuned to be attractive, and the atoms
are cooled to sufficiently low temperatures, the components will form
pairs via the Cooper instability. Due to this pair formation the Fermi
gas becomes a Bardeen-Cooper-Schrieffer (BCS) superfluid as was
envisaged in Refs.~\cite{Stoof96, Houbiers97}.

The response of the superfluid to rotation can be investigated by
rotating the trapping potential with a certain frequency $\Omega$. For
a non-rotating trap the entire gas will form a superfluid without
vortices. Let us now imagine increasing the rotation frequency at zero
temperature. Up to a certain critical rotation frequency, the
superfluid will stay in the vortex-free state carrying zero angular
momentum. For low temperatures the angular momentum will be quenched,
as has also been observed experimentally~\cite{AngularMomentumQuenching}.
Above the critical frequency, angular momentum will be inserted in the
gas by either unpairing the fermions near the edges of the gas, by
formation of vortices, or by the combination of both effects. The goal
of this paper is to compute the rotation frequencies at which these
transitions take place.

Besides the experimental investigations \cite{Zwierlein05,
  Zwierlein06, Zwierlein06b, Schunck07, AngularMomentumQuenching},
various theoretical studies of rotating two-component Fermi gases have
been performed (for a review see e.g.~Ref.~\cite{Giorgini08}).  The
profile of a single vortex was analyzed in this context for the first
time in Ref.~\cite{Rodriguez01} using a Ginzburg-Landau approach and
in Ref.~\cite{Nygaard03} by solving the Bogoliubov-de Gennes equation.
In Ref.~\cite{Bulgac03} it was concluded that a single vortex can
induce a sizable density depletion at its core. This was also found to
be the case for a vortex lattice~\cite{Feder04}. Density depletion is
important, since it allows the vortices to be detected experimentally
\cite{Zwierlein05}.  The vortex profile was investigated in a
population imbalanced gas in Ref.~\cite{Takahashi06} and in situation
in which the two components have unequal mass in
Ref.~\cite{Iskin08}. Real-time dynamics of vortices has been studied
in Ref.~\cite{Bulgac11}.

Vortex lattices in two-component Fermi gases have been examined in
several ways in Refs.~\cite{Feder04, Botelho06, Tonini08}. At high
rotation frequencies a completely unpaired phase is preferred over a
vortex lattice. The critical rotation frequency corresponding to this
transition was computed in Refs.~\cite{Zhai06, Veillette06}.  A vortex
lattice can also be destroyed by heating the gas. The corresponding
critical temperature was computed in Refs.~\cite{Feder04, Moller07}.
When the number of trapped components is unequal, a vortex lattice
might be formed within the Fulde-Ferrell-Larkin-Ovchinnikov (FFLO)
phase.  Its melting temperature was investigated in Ref.~\cite{Shim06}
and its upper critical rotation frequency in Ref.~\cite{Kulic09}.

The first analysis of the critical frequency $\Omega_c$ for vortex
formation in a two-component Fermi gas was performed in
Ref.~\cite{Bruun01}.  To obtain $\Omega_c$ the Helmholtz free energy
difference $\Delta F$ between a vortex with unit angular momentum
located at the center of the trap and the vortex-free superfluid was
estimated at $\Omega = 0$. For a cylindrically symmetric infinite well
as the trapping potential, $\Delta F$ was obtained by solving the
Bogolibuov-de Gennes equation in Refs.~\cite{Nygaard03, Nygaard04}.
The free energy difference arises from the loss of condensation energy
at the vortex core, the kinetic energy of fermions circulating around
the vortex core, and the energy needed to expand the cloud to
accommodate the excess particles removed from the vortex core (the
latter effect was not considered in Ref.~\cite{Bruun01}, but was taken
into account in Refs.~\cite{Nygaard03, Nygaard04}). When rotating the
trap, the free energy decreases with $\Omega L_z$, where $L_z$ is the
angular momentum contained in the gas. For the vortex-free superfluid
$L_z = 0$, while for the vortex with unit angular momentum $L_z =
N\hbar /2 $, with $N$ the number of trapped particles. These estimates
are correct if rotating the trap does not cause unpairing near the
edges of the gas. The critical rotation frequency in this case is
$\Omega_c = 2\Delta F / (N \hbar)$.  This frequency is similar to the
lower critical magnetic field in type-II
superconductors~\cite{Bardeen69}.

Fermions confined in a rotating trap can unpair near the
edges~\cite{Bausmerth08, Urban08, Iskin09}. This effect was not
considered in Refs.~\cite{Bruun01, Nygaard03,Nygaard04}. In this paper
we will take into account this possibility in order to obtain a more
reliable value of $\Omega_c$. We will make detailed study of how
$\Omega_c$ varies with the interaction strength and the number of
trapped particles. Furthermore we will compute the critical rotation
frequency for unpairing.  We consider a cylindrically symmetric 
harmonic trapping potential.  Our calculations are carried out in the
Hartree-Fock-Bogoliubov approximation. Therefore, we expect that our
results are reliable only in the weak coupling limit.

Let us finally remark that vortices have also been observed in
rotating Bose-Einstein condensates (BEC) of bosonic atoms
\cite{BECvortices} and of bosonic dimers composed of fermionic atoms
\cite{Zwierlein05} (for a review see e.g.~Ref.~\cite{Fetter09}). The
behavior of vortices through the BEC-BCS crossover was investigated
experimentally in Ref.~\cite{Zwierlein05}. In this article we only
discuss the BCS regime. A computation of the critical rotation
frequency for vortex formation in a BEC is discussed in
Ref.~\cite{Stringari99}.  Theoretical studies of behavior of vortices
through the BEC-BCS crossover have been performed in
Refs.~\cite{Kawaguchi04, Chien06, Sensarma06}.

This article is organized as follows. In Sec.~\ref{sec:setup} we will
introduce the action from which one can derive the properties of the
Fermi gas. To achieve this in practice we will adopt the two-particle
irreducible (2PI) effective action, which we explain in
Sec.~\ref{sec:2pi}. From the 2PI effective action one obtains the
Dyson-Schwinger equation which is the main equation we have to
solve. This can be achieved by finding the solution of the
Bogoliubov-de Gennes equation, which is explained in
Sec.~\ref{sec:bdg}. The numerical methods by which we have solved the
Bogoliubov-de-Gennes and the Dyson-Schwinger equation are discussed in
Secs.~\ref{sec:sbdg} and \ref{sec:ds} respectively.  The reader who is
not interested in the details of the calculation can immediately go to
Sec.~\ref{sec:results} where we present the results.  The phase
diagrams presented in Figs.~\ref{fig:PD1000} and \ref{fig:PD200} are
our main results. We draw our conclusions in
Sec.~\ref{sec:conclusions}.  Several details are relegated to the
appendices. In Appendix~\ref{app:2pi} we review the 2PI effective
action. A derivation of the Bogoliubov-de Gennes equation is presented
in Appendix~\ref{app:BdG}. To solve the Bogoliubov-de Gennes equation
numerically, we will use a basis based on Maxwell polynomials. We
discuss the computation of the quadrature weights and nodes of these
polynomials in Appendix~\ref{app:maxpoly}. In Appendix~\ref{app:barh} we
derive the representation of the single particle Hamiltonian in the
basis we will employ.

\section{Setup}
\label{sec:setup}
Let us consider a two-component Fermi gas in which $s$-wave
interactions are dominant and label its components by $\alpha =\uparrow,
\downarrow$.  Typically in experimental realizations the higher
partial waves can be neglected and the superfluidity is driven by
attractive $s$-wave interactions. The interactions among the same
components can be neglected since the Pauli principle admits $s$-wave
interactions only between the different species. We will denote the
$s$-wave interaction potential as $V(\vec x)$ and specify it
below. Under these assumptions the interacting Fermi gas is described
by the following action (see e.g.\ Ref.~\cite{Stoof}), $S =
S_{\mathrm{kin}} + S_{\mathrm{int}}$ where
\begin{gather}
S_\mathrm{kin} = 
\sum_{\alpha=\uparrow, \downarrow} 
\int \rmd X
\psi_{\alpha}^*(X) 
\left[
\hbar \frac{\partial}{\partial \tau}
+
H(\Omega) - \mu_\alpha \right] \psi_\alpha(X)
,
\label{eq:action}
\\
\begin{split}
S_{\mathrm{int}} 
& = 
\frac{1}{2}
\sum_{\alpha=\uparrow, \downarrow}
\int \rmd X
\int \rmd Y \,
\psi^*_\alpha(X_+)
\psi^*_{-\alpha}(Y_+)
\\
& \quad \quad \quad \quad \times
V(\vec x- \vec y) \delta(\tau_x - \tau_y)
\psi_{-\alpha}(Y)
\psi_{\alpha}(X)
\label{eq:actioninteraction}.
\end{split}
\end{gather}
Here $\psi_\alpha(X)$ is the (path-integral) quantum field
corresponding to the $\alpha$ component, and $X = (\vec x, \tau_x)$.
We write integration over spatial coordinates and imaginary time
$\tau$ as $\int \rmd X \equiv \int_0^{\hbar \beta} \rmd \tau_x \int
\rmd^3 x$ with the inverse temperature $\beta = 1/(k_B T)$.  Here and
in the rest of the article $X_\pm = (\vec x, \tau_x \pm \eta)$ and
$\eta$ is an infinitesimal small positive number. We have inserted
$X_+$ and $Y_+$ in Eq.~(\ref{eq:actioninteraction}) in order to
maintain the correct ordering of the fields in the path-integral.
We will achieve this in a different way for the kinetic term and
explain this at the end of Appendix~\ref{app:BdG}.  Furthermore
$-\uparrow$ is equivalent to $\downarrow$, $\mu_\alpha$ denotes the
chemical potential, and $H(\Omega)$ is the single-particle
Hamiltonian. We will assume that the particles are trapped in a
potential that is rotating in the $x$-$y$ plane with angular frequency
$\Omega$. It is then convenient to perform the calculation in the
rotating frame.  The single-particle Hamiltonian $H(\Omega)$ in the
rotating frame reads (see e.g.\ Ref.~\cite{Stoof})
\begin{equation}
 H(\Omega) = \frac{\vec p^2}{2M} + U(\vec x) 
- \Omega L_z,
\label{eq:cyltrap}
\end{equation}
where $M$ is the fermion mass and $L_z = x p_y - y p_x$ is the $z$-component
of the angular momentum.  The trapping potential $U(\vec x)$ realized
in experiments is typically harmonic. In this paper we will study a
cylindrically shaped trap given by the potential
\begin{equation}
  U(\vec x) = \frac{1}{2} M \omega^2 (x^2 + y^2),
\end{equation}
which implies confinement in the $x$-$y$ plane, and infinite extension
in the $z$-direction. In an experiment this regime can be reached by
choosing the trapping frequency in the $z$-direction much smaller than
in the $x$-$y$-direction.  In cylindrical coordinates, $\vec x = (\rho
\cos \phi, \rho \sin \phi, z)$, the single-particle Hamiltonian reads
\begin{equation}
\begin{split}
  H(\Omega) & = \frac{\hbar^2}{2M} \left( -
    \frac{\mathrm{d}^2}{\mathrm{d} \rho^2}
    -
    \frac{1}{\rho} \frac{\mathrm{d}}{\mathrm{d} \rho}
    + \frac{L_z^2}{\hbar^2 \rho^2}
  - \frac{\mathrm{d}^2}{\mathrm{d} z^2}
  \right)
\\
& \quad
  + \frac{1}{2} M \omega^2 \rho^2 
  - \Omega L_z,
\label{eq:Hsp}
\end{split}
\end{equation}
where $L_z = -\rmi \hbar \partial/\partial \phi$.  The normalized
eigenfunctions $\psi^0_{nmp_z}(\vec x)$ of this Hamiltonian can be
written as a product of three functions,
\begin{equation}
\psi^0_{nmp_z}(\vec x)
 = \frac{1}{\sqrt{L}} R_{nm}(\rho) f_m(\phi) \rme^{\rmi p_z z},
\label{eq:efhsp}
\end{equation} 
with the radial quantum number $n=0,1,\ldots$, the angular momentum
quantum number $m \in \mathbb{Z}$, and the momentum in the
$z$-direction $p_z = 2 \pi \hbar n_z / L$ with $n_z \in
\mathbb{Z}$. The constant $L$ denotes the length of the system in the
$z$-direction. In this article we will consider the limit $L
\rightarrow \infty$. Then $\frac{1}{L} \sum_{n_z} = \int \rmd p_z /
(2\pi \hbar)$. The function $f_m(\phi)$ is a normalized eigenfunction
of $L_z$ and is given explicitly by
\begin{equation}
 f_m(\phi) = \frac{1}{\sqrt{2\pi}} e^{ \mathrm{i} m \phi}.
\end{equation}
The radial eigenfunctions are given by
\begin{equation}
R_{nm}(\rho) = c_{nm} L_{n}^{\vert m \vert}( \bar \rho^2) \bar \rho^{\vert m\vert}
e^{-\bar \rho^2/2},
\label{eq:rwhsp}
\end{equation}
where $L_n^a(x)$ denotes the generalized Laguerre polynomial which has
degree $n$. Furthermore $\bar \rho = \rho/ \lambda$
with the harmonic oscillator length $\lambda = (\hbar / M
\omega)^{1/2}$.  Normalization gives $c_{nm}^2 = 2n! / [\lambda^2
(n+\vert m\vert)!]$. The energy spectrum of $H(\Omega)$ is given by
\begin{equation}
\epsilon^0_{nmp_z} = \hbar \omega 
\left( 1 + 2 n + \vert m \vert\right) 
- \hbar \Omega m 
+ \frac{p_z^2}{2M}.
\end{equation}

At low enough temperatures and densities, the typical wavelength of
the particles will be much longer than the range of the interaction
potential. In that case the detailed structure of the potential is
unimportant and the only relevant interaction parameter is the
scattering length $a$. To perform a calculation in this situation, one
can just choose the most convenient potential that has scattering
length $a$. Following Ref.~\cite{Bruun99} we will use the Huang-Yang
potential \cite{Huang57}
\begin{equation}
 V(\vec r) = g \delta(\vec r) \frac{\partial}{\partial r} r,
 \label{eq:hypotential}
\end{equation}
where the coupling constant $g = 4 \pi a \hbar^2/ M$. Pairing between
fermions requires attractive interactions, that is $a < 0$ and hence
$g < 0$. Note that the Huang-Yang potential is not equivalent to an
ordinary $\delta$-function potential $g \delta(\vec r)$, because the
derivative operator also acts on the fields in
Eq.~(\ref{eq:actioninteraction}). The advantage of the Huang-Yang
potential is that all relevant physical quantities become
automatically convergent \cite{Bruun99}. Furthermore, the coupling
constant does not have to be renormalized so that the foregoing
relation between the scattering length and the coupling constant
always holds \cite{Bruun99}.  In the case of an ordinary
$\delta$-function potential one will encounter divergences.  This will
require a regularization prescription and a renormalization of the
coupling constant.

In order to perform calculations it is convenient to rewrite the
action in the Nambu-Gor'kov basis.  For that purpose we introduce the
Nambu-Gor'kov fields
\begin{equation}
\Psi(X) = \left
  ( \begin{array}{c} 
\psi_{\uparrow}(X) 
\\
\psi^*_{\downarrow}(X)
\end{array}
\right ),
\end{equation}
and write the kinetic part of the action, Eq.~(\ref{eq:action}), as
\begin{equation}
 S_{\mathrm{kin}} = -\hbar 
\int \rmd X \int \rmd X' \,
\Psi^\dagger(X) G_0^{-1}(X, X')
\Psi(X'),
\end{equation}
where the bare inverse Nambu-Gor'kov propagator reads
\begin{multline}
G_0^{-1}(X, X')
=
\\
- \frac{1}{\hbar}
 \left(
\begin{array}{cc}
\hbar \frac{\partial}{\partial \tau}
 +
 H(\Omega) - \mu_\uparrow
&
0
\\
0
&
\hbar \frac{\partial}{\partial \tau}
- H(\Omega)^* + \mu_\downarrow
\end{array}
\right)
\\
\times
 \delta(X - X').
\label{eq:bareinvnambugorkov}
\end{multline}
Here we  used the fact that $H(-\Omega) = H(\Omega)^*$.  In the
Nambu-Gor'kov basis the interaction part of the action becomes 
\begin{equation}
\begin{split}
 S_{\mathrm{int}} &
= -  \sum_{\alpha = \pm}
\int \rmd X \rmd Y \rmd X' \rmd Y' \,
\Psi^\dagger(X') \sigma_{\alpha} \Psi(X)
\\ 
& \quad \quad \quad \times
\Psi^\dagger(Y') \sigma_{-\alpha} \Psi(Y) 
\mathcal{V}_\alpha(X, Y; X', Y'),
\label{eq:NGSint}
\end{split}
\end{equation}
where $\sigma_+ = \mathrm{diag}(1, 0)$ and $\sigma_- =
\mathrm{diag}(0, 1)$. The potential $\mathcal{V}_\alpha$ is given by
\begin{multline}
\mathcal{V}_\alpha(X, Y; X', Y') 
= 
\frac{1}{4}
\Bigl[
(1+\alpha)
\delta(\tau_{x'} - \tau_y) 
V(\vec x' - \vec y)
\\
+
(1 \!-\! \alpha)
\delta(\tau_x \!- \!\tau_{y'}) 
V(\vec x - \vec y') 
\Bigr] 
\delta(X_\alpha \!-\! X')\delta(Y_{-\alpha}\!-\!Y').
\label{eq:calV}
\end{multline}
Because the Huang-Yang potential contains a
derivative operator, we had to introduce several $\delta$-functions 
in order to  separate the potential operator from the quantum fields.

\section{2PI effective action}
\label{sec:2pi}

To study the interacting Fermi gas we will compute the resummed
propagator $G_{ij}(X,X') = -\langle \Psi_i(X) \Psi_j^\dagger(X')
\rangle$ and the grand potential $\Phi_G$ by using the two-particle
irreducible (2PI) effective action \cite{2PI}. Readers who are not
interested in the details of this formalism can immediately continue
with Sec.~\ref{sec:bdg} where we discuss the Bogoliubov-de Gennes
equation which follows from the 2PI effective action.

The 2PI effective action is also known as the
Cornwall-Jackiw-Tomboulis formalism and is equivalent to the
Luttinger-Ward functional approach~\cite{LW}. It has been applied
previously to investigate pairing in atomic gases \cite{Haussmann07}
and in quark matter \cite{CSC2PI}.
 
The main advantage of the 2PI effective action is that it is a
functional method which generates the resummed Nambu-Gor'kov
propagator and the corresponding grand potential. Here, we will
truncate the 2PI effective action at order $g$, which is equivalent to
the Hartree-Fock-Bogoliubov approximation. This leads to the well known
Bogoliubov-de Gennes equation. Another advantage of the 2PI method is
that any truncation can be systematically improved straightforwardly
by taking into account higher order diagrams. This is necessary when
extending our results to a strongly coupled Fermi gas.

\begin{figure}[t]
\begin{tabular}{ll}
\includegraphics[width=2.8cm]{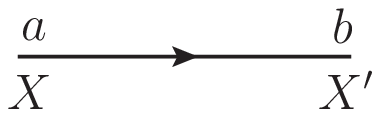}
& \raisebox{0.5cm}{$= G_{ab}(X, X') $} \\
\includegraphics[width=2.8cm]{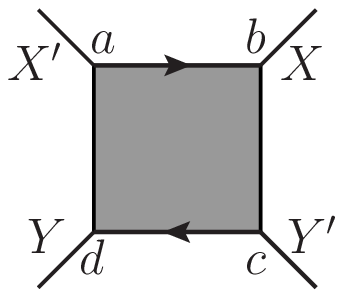} &
\raisebox{1cm}{$\displaystyle{ = \sum_{\alpha=\pm} 
 (\sigma_{\alpha})_{ab} (\sigma_{-\alpha})_{cd} 
\mathcal{V}_\alpha(X, Y; X', Y')}$}
\end{tabular}
\caption{Feynman rules for the propagator and the vertex.  When
  connecting propagators to the vertex, the arrows of the propagators
  have to point in the same direction as the arrows in the vertex.
  For each closed fermion loop one should include a
  factor $-1$. 
  \label{fig:feynmanrules}}
\end{figure}

\begin{figure}[tb]
\includegraphics[width=8cm]{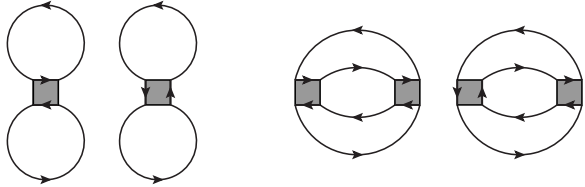}
\caption{All 2PI diagrams contributing to $\Gamma_2[G]$ at order $g$ and order $g^2$.
Note the direction of the arrows.
\label{fig:gamma2}}
\end{figure}

The 2PI effective action reads (see Appendix~\ref{app:2pi} for details) 
\cite{2PI}
\begin{equation}
  \Gamma[G] = -\mathrm{Tr} \log G^{-1} - \mathrm{Tr} 
  \left( G_0^{-1} G - 1 \right)
  + \Gamma_2[G],
\end{equation}
where $\Gamma_2[G]$ is the sum of all 2PI diagrams generated from
$S_{\mathrm{int}}$ with propagators $G$. The interaction vertex can be
directly read off from Eq.~(\ref{eq:NGSint}). We have displayed the
Feynman rules for computing the 2PI effective action in
Fig.~\ref{fig:feynmanrules}. All diagrams contributing to $\Gamma_2[G]$
up to order $g^2$ are displayed in Fig.~\ref{fig:gamma2}.


By minimizing $\Gamma[G]$ with respect to $G$ one obtains the
Dyson-Schwinger equation
\begin{equation}
 G^{-1} = G_0^{-1} - \Sigma[G], 
 \label{eq:ds}
\end{equation}
where the 1PI self-energy is $\Sigma[G] = \delta \Gamma_2[G] / \delta G$.
The grand potential $\Phi_G = - \frac{1}{\beta} \log Z$ is equal
to the minimum value of $\Gamma[G]/\beta$ and reads
\begin{equation}
  \Phi_G = - \frac{1}{\beta} \mathrm{Tr} \log G^{-1}
 - \frac{1}{\beta} \mathrm{Tr}\left( \Sigma[G] G \right) + 
\frac{1}{\beta}
\Gamma_2[G],
\label{eq:thermopotA}
\end{equation}
where $G$ is now the solution of the Dyson-Schwinger equation,
Eq.~(\ref{eq:ds}).

In order to perform a practical calculation, one has to truncate
$\Gamma_2[G]$ at some order. In this paper we will take into account
the contributions of order $g$ to $\Gamma_2[G]$, which are represented
by the first two diagrams in Fig.~\ref{fig:gamma2}. Such truncation
leads to the Hartree-Fock-Bogoliubov approximation. This approximation
is expected to give accurate results for small coupling $g$. In general 
the Hartree-Fock-Bogoliubov approximation  can
also be applied to strongly correlated systems if the interaction
kernel is renormalized appropriately. Such renormalization entails a
resummation of ladder diagrams in the Lippman-Schwinger
equation. Since we are interested in the weakly interacting BCS limit
there is no need to do this.

Applying the Feynman rules to the first two diagrams in
Fig.~\ref{fig:gamma2} we find that to order $g$, $\Gamma_2[G]$ is given by
\begin{equation}
\begin{split}
\Gamma_2[G] 
&
=
-\frac{1}{\hbar} \sum_{\alpha = \pm}
\int \rmd X \rmd Y \rmd X' \rmd Y' \,
\mathcal{V}_{\alpha}(X, Y; X', Y')
\\
&\quad \quad \quad \quad \quad
\times
\mathrm{tr} \left[G(X, X') \sigma_{\alpha} \right]  
\mathrm{tr} \left[G(Y, Y') \sigma_{-\alpha} \right] 
\\
& \quad
+ \frac{1}{\hbar} \sum_{\alpha = \pm} 
\int \rmd X \rmd Y \rmd X' \rmd Y' \,
\mathcal{V}_{\alpha}(X, Y; X', Y')
\\
&
\quad \quad \quad \quad \quad
\times
\mathrm{tr}
\left[ 
G(X, Y') \sigma_\alpha G(Y, X') \sigma_{-\alpha} 
\right].
\label{eq:gamma2B}
\end{split}
\end{equation}
The relative minus sign arises because the first diagram in
Fig.~\ref{fig:gamma2} contains two closed loops whereas the second
diagram contains only one closed loop when following the arrows. 

The self-energy $\Sigma[G]$ can be obtained by differentiating
Eq.~(\ref{eq:gamma2B}) with respect to $G$. This yields
\begin{equation}
\begin{split}
 \Sigma[G](X, X') 
& =
- \frac{2}{\hbar}
\sum_{\alpha=\pm} \sigma_a
\int \rmd Y \rmd Y' \,
\mathrm{tr}
\left[G(Y, Y') \sigma_{-\alpha} \right]
\\
& 
\quad \quad \quad
\times
\mathcal{V}_{\alpha}(X', Y; X, Y')
\\
& \quad +
\frac{2}{\hbar}
\sum_{\alpha=\pm} 
\int \rmd Y \rmd Y' \,
\sigma_{\alpha} G(Y, Y') \sigma_{-\alpha}
\\
&
\quad \quad \quad
\times
\mathcal{V}_{\alpha}(X', Y; Y', X).
\label{eq:sigmaB}
\end{split}
\end{equation}
The first term is the Hartree self-energy, the second one is the 
pairing contribution.

The last two equations can be simplified by computing the traces and
inserting the explicit expressions for $\mathcal{V}$ and $V(\vec r)$.
One has then to act the Huang-Yang potential on the propagators.
It can be shown \cite{Bruun99} that the diagonal components of $G(X,
X')$ are finite in the limit $r \rightarrow 0$ where $r = \vert\vec x
- \vec x' \vert$. For these components the Huang-Yang potential
acts as an ordinary $\delta$-function potential. We will use this fact
to simplify the expressions involving the diagonal components of $G$.
On the other hand the off-diagonal components of $G$ will have in
general a singularity proportional to $1/r$ when $r \rightarrow 0$
\cite{Bruun99}. In such situations the full form of the Huang-Yang
potential needs to be taken into account. 

For later convenience we will now define the pairing field as 
\begin{equation}
 \Delta(\vec x) \equiv \int \rmd^3 x'\,
V(\vec x - \vec x') G_{\uparrow \downarrow}(\vec x, \tau; 
\vec x', \tau).
\label{eq:delta_VG}
\end{equation}
Following Ref.~\cite{Bruun99} we will split the off-diagonal component
of the Nambu-Gor'kov propagator into a singular and regular part
\begin{equation}
  \lim_{\vec x' \rightarrow \vec x} 
  G_{\uparrow \downarrow}(\vec x, \tau; \vec x', \tau)
  = \frac{C}{r} +  G^{\mathrm{reg}}
_{\uparrow \downarrow}(X, X),
\end{equation}
where $C$ is some constant and the superscript reg indicates the
regular part of $G$. By inserting the Huang-Yang potential in 
Eq.~(\ref{eq:delta_VG}) one can see that the pairing field does not
contain any singularity~\cite{Bruun99}
\begin{equation}
 \Delta(\vec x) = g G^{\mathrm{reg}}
_{\uparrow \downarrow}(X, X).
\label{eq:delta}
\end{equation}

The diagonal components of the propagator can be expressed in terms of
number densities. To find the relation between the propagator and the 
densities  we use the definition of the propagator in terms of field
operators
\begin{equation}
\begin{split}
G_{ij}(X, X') & 
= - \langle T_\tau \hat \Psi_i(X) \hat
\Psi_j^\dagger(X') \rangle 
\\
& =  \theta(\tau' - \tau)  \langle \hat
\Psi_j^\dagger(X')\hat\Psi_i(X) \rangle
\\
& \quad - \theta(\tau - \tau') 
\langle
\hat \Psi_i(X) \hat \Psi_j^\dagger(X') \rangle
\label{eq:propfieldoperators}
,
\end{split}
\end{equation}
here $T_\tau$ denotes time ordering in imaginary time. Hence the
number densities of the two species are related to $G$ as 
\begin{eqnarray}
n_\uparrow(\vec x) &=& \langle 
\hat \psi_{\uparrow}^\dagger(\vec x, \tau)
\hat \psi_{\uparrow}(\vec x, \tau) \rangle 
=
G_{\uparrow \uparrow}(X, X_+),
\label{eq:densup}
\\
n_\downarrow(\vec x) &=& \langle 
\hat \psi_{\downarrow}^\dagger(\vec x, \tau)
\hat \psi_{\downarrow}(\vec x, \tau)\rangle =
-G_{\downarrow \downarrow}(X, X_-).
\label{eq:densdown}
\end{eqnarray}

From Eq.~(\ref{eq:propfieldoperators}) it follows that $G_{\uparrow
  \downarrow}(X, X') = \langle \hat \psi_{\downarrow}(X') \hat
\psi_{\uparrow}(X) \rangle$ and $G_{\downarrow \uparrow}(X, X') =
\langle \hat \psi^{\dagger}_{\uparrow}(X') \hat
\psi^{\dagger}_{\downarrow}(X) \rangle$. This can be used to show that
$G_{\downarrow \uparrow}(X, X') = G_{\uparrow \downarrow}(X',
X)^*$ which implies that
\begin{equation}
\begin{split}
 \Delta(\vec x)^* 
& 
= \int \rmd^3 x' \,
V(\vec x -\vec x') G_{\downarrow \uparrow}
(\vec x, \tau; \vec x', \tau)
\\
& =
 g G^{\mathrm{reg}}_{\uparrow \downarrow}
(X, X)^*
.
\end{split}
\end{equation}
Here we used the fact that the singular part of $G_{\uparrow
  \downarrow}(X',X)^*$ is a function of $\vert \vec x - \vec x'\vert
$, so that $\vec x$ and $\vec x'$ could be interchanged without 
changing the result.

We can now use the above definitions to simplify
Eqs.~(\ref{eq:gamma2B}) and (\ref{eq:sigmaB}). After computing the
traces we find that to order $g$, $\Gamma_2[G]$ is given by
\begin{equation}
\begin{split}
\frac{1}{\beta} \Gamma_2[G] &
= g \int \rmd^3 x \, 
n_{\uparrow}(\vec x) n_{\downarrow}(\vec x )
\\
&
\quad + \int \rmd^3 x \,
G_{\uparrow \downarrow}(X, X)^*
\Delta(\vec x)
\label{eq:gamma2C}.
\end{split}
\end{equation}
The 1PI self-energy to order $g$ is
\begin{equation}
\Sigma[G](X,X')
=
\frac{1}{\hbar} 
\left(
\begin{array}{cc}
g n_{\downarrow}(\vec x)
&
\Delta(\vec x )
\\
\Delta^*(\vec x)
&
-g n_{\uparrow}(\vec x)
\end{array}
\right)
\delta(X-X').
\label{eq:sigmaC}
\end{equation}
It follows that $\mathrm{Tr}(\Sigma[G]G) = 2 \Gamma_2[G] + O(g^2)$
so that the grand potential $\Phi_G$, Eq.~(\ref{eq:thermopotA}),
to order $g$ becomes
\begin{equation}
\begin{split}
\Phi_G & = -\frac{1}{\beta} \mathrm{Tr} \log G^{-1}
- 
g \int \rmd^3 x \, 
n_{\uparrow}(\vec x) n_{\downarrow}(\vec x )
\\
& 
\quad - 
\int \rmd^3 x \,
G_{\uparrow \downarrow}(X, X)^*
\Delta(\vec x).
\label{eq:thermopotB}
\end{split}
\end{equation}

Since the particle number in the trap is fixed we solve the equations
for the chemical potentials to obtain the desired number of
fermions in each hyperfine state.  The appropriate thermodynamic
potential is then the Helmholtz free energy, which reads
\begin{equation}
 F = \Phi_G + \mu_\uparrow N_\uparrow + \mu_\downarrow N_\downarrow,
\label{eq:FHelmholtz}
\end{equation}
where $N_{\uparrow, \downarrow}$ denote the total number of particles
of a particular species. Since we consider a cylindrically shaped
trap, $F$ and $N_{\uparrow, \downarrow}$ are proportional to the
length of the trap in the $z$-direction $L$.  Since $L$ is taken to be
infinite is convenient to consider instead the free energy and
particle number per unit of the harmonic oscillator length $\lambda$
in the $z$-direction. For this reason we define
\begin{equation}
\mathcal{F} = \frac{F}{L / \lambda}, \;\;\;\;\;
\mathcal{N}_{\uparrow, \downarrow} 
= \frac{N_{\uparrow, \downarrow}}{L / \lambda}.
\end{equation}
Furthermore we will define $\mathcal{N}$ to be the total number of
particles per unit of length in the $z$-direction, $\mathcal{N} =
\mathcal{N}_\uparrow + \mathcal{N}_\downarrow$.

\section{Bogoliubov-de Gennes equation}
\label{sec:bdg}
To proceed, we will insert the explicit expression for the 1PI
self-energy, Eq.~(\ref{eq:sigmaC}), into Eq.~(\ref{eq:ds}).  This
yields the Dyson-Schwinger equation for the Nambu-Gor'kov propagator,
\begin{equation}
G^{-1}(X, X')
=
- \frac{1}{\hbar} \left( \hbar \frac{\partial}{\partial \tau} + 
\mathcal{H} \right) \delta(X-X'),
\label{eq:dsB}
\end{equation}
with
\begin{multline}
\mathcal{H} = \\
\left(
\begin{array}{cc}
 H(\Omega) \!-\! \mu_\uparrow \! +\! g n_{\downarrow}(\vec x)
&
\Delta(\vec x )
\\
\Delta^*(\vec x)
&
- H(\Omega)^* \!+\! \mu_\downarrow \!-\! g n_{\uparrow}(\vec x) 
\end{array}
\right).
\label{eq:hbdg}
\end{multline}
To solve the Dyson-Schwinger equation, one first inverts both the left- and
right-hand sides of Eq.~(\ref{eq:dsB}).  As explained in detail in
Appendix~\ref{app:BdG} this can be achieved by solving the
Bogoliubov-de Gennes equation \cite{Gennes64}
\begin{equation}
\mathcal{H}
\left(
\begin{array}{c}
u_i(\vec x)
\\
v_i(\vec x)
\end{array}
\right)
= 
E_i
\left(
\begin{array}{c}
u_i(\vec x)
\\
v_i(\vec x)
\end{array}
\right).
\label{eq:BdGB}
\end{equation}
The functions $u_i(\vec x)$ and $v_i(\vec x)$ have to be normalized as
$\int \rmd^3 x \, \left[ \vert u_i(\vec x) \vert^2 + \vert v_i(\vec x
  ) \vert^2 \right] = 1$. Using the explicit expression of $G$,
Eq.~(\ref{eq:G1}), and  Eqs.~(\ref{eq:densup}),
(\ref{eq:densdown}) one can now read off the densities 
\begin{eqnarray}
 n_{\uparrow}(\vec x) &=& \sum_i f(E_i) \vert u_i(\vec x) \vert^2,
\label{eq:densupA}
 \\ 
 n_{\downarrow}(\vec x) &=& \sum_i f(-E_i) \vert v_i(\vec x) \vert^2,
\label{eq:densdownB}
\end{eqnarray}
where $f(E) = [\exp(\beta E) + 1]^{-1}$ is the Fermi-Dirac
distribution function. 

As follows from Eqs.~(\ref{eq:delta}) and (\ref{eq:G1})
to obtain $\Delta(\vec x)$ we need to extract the regular part of
the propagator
\begin{equation}
G_{\uparrow \downarrow}(\vec x, \tau; \vec x', \tau) 
= \sum_i f(E_i) u_i(\vec x) v_i^*(\vec x'),
\label{eq:gupdown}
\end{equation}
in the limit $\vec x' \rightarrow \vec x$. To do so
we will use the method proposed in Ref.~\cite{Bruun99} with
the improvements suggested in Refs.~\cite{Bulgac02,Grasso03}.

The sum over all modes in Eq.~(\ref{eq:gupdown}) is logarithmically
divergent for $\vec x = \vec x'$. The singular part of $G_{\uparrow
  \downarrow}$ arises from the modes in the integrand with large
negative energy.  To obtain the regular part in the limit $\vec x'
\rightarrow \vec x$ we will first subtract this large energy
contribution. For this purpose we define
\begin{equation}
\nu_c(\vec x) = \sum_{\vert E_i \vert < E_c} f(E_i) u_i(\vec x) v_i^*(\vec x),
\label{eq:nuc}
\end{equation}
where $E_c$ denotes an energy cutoff introduced to regulate the
logarithmic divergence in $\nu_c(\vec x)$. The part of $\nu_c(\vec x)$
dominated by the modes with large negative energies can be
approximated as \cite{Bruun99, Bulgac02, Grasso03} $\nu_{HE}(\vec x) =
- \Delta(\vec x) K(\vec x, \vec x; E_c)$ with
\begin{equation}
K(\vec x, \vec x'; E_c) = 
\sum_{E_s < \epsilon_i < E_c} \frac{\psi_i(\vec x) \psi^*_i(\vec x')}
{2 \epsilon_i }, 
\label{eq:Kx}
\end{equation}
where here and below $\psi_i(\vec x)$ and
$\epsilon_i$ denote the eigenvectors and eigenvalues of the
Hartree-Fock Hamiltonian
\begin{equation}
H_{\mathrm{HF}} = H(\Omega = 0) - \mu + g n(\vec x).
\label{eq:HHF}
\end{equation}
Here $\mu = (\mu_\uparrow + \mu_\downarrow)/2$ and $n(\vec x) =
(n_{\uparrow}(\vec x) + n_{\downarrow}(\vec x))/2$.  The low-energy
cut-off $E_s$ in Eq.~(\ref{eq:Kx}) is arbitrary, except that it should
be chosen positive in order to avoid singularities arising from the
Fermi surface.  As an alternative to introducing a low-energy cutoff,
one can add a small imaginary part to $\epsilon_i$ as done in
Refs.~\cite{Bruun99, Bulgac02, Grasso03}. In that case the integrand
of $K(\vec x, \vec x'; E_c)$ has a peak near the Fermi surface,
which makes the numerical integration over $p_z$ difficult.  One can
reduce this peak by increasing the magnitude of the imaginary
part. However, that will worsen the large negative energy
approximation of $\nu_c(\vec x)$. The low-energy cutoff which we apply
here does not suffer from these problems.

Let us next define $\nu_s(\vec x) \equiv \nu_c(\vec x) - \nu_{HE}(\vec
x)$.  Because $\nu_{HE}(\vec x)$ contains the logarithmic divergent
part of $\nu_c(\vec x)$, the difference $\nu_s(\vec x)$ is finite, and
hence converges for large enough $E_c$.  There is some freedom in
choosing $\nu_{HE}(\vec x)$. For example, we could have left out the
$g n(x)$ term in the Hartree-Fock Hamiltonian as in
Refs.~\cite{Bruun99, Bulgac02}. However, by including this term,
$\nu_{HE}(\vec x)$ approximates $\nu_c(\vec x)$ much better, which
implies that a much smaller value of $E_c$ is sufficient to compute
$\nu_s(\vec x)$ accurately \cite{Grasso03}.

Summarizing the discussion above, we found that in the limit $\vec x'
\rightarrow \vec x$, we can write $ G_{\uparrow \downarrow}(\vec x,
\tau; \vec x', \tau) = \nu_s(\vec x) - \Delta(\vec x) K(\vec x, \vec
x'; \infty)$.  Following Refs.~\cite{Bulgac02, Grasso03} the singular
part can now be obtained by making use of the Thomas-Fermi
approximation. In the limit $\vec x' \rightarrow \vec x$ one finds
that
\begin{multline}
 K(\vec x, \vec x'; \infty)
 =
 K(\vec x, \vec x; E_{c'})
\\
 -
\frac{1}{2} \int_{k < k_{c'}(\rho)} \frac{\rmd^3 k}{(2 \pi)^3}
\frac{1}
{ \frac{\hbar^2 \vec k^2}{2M} + \frac{1}{2} M \omega^2 \rho^2 
- \mu + gn(\rho) + \rmi \gamma}
\\
+ \frac{1}{2}
\int \frac{\rmd^3 k}{(2 \pi)^3}
\frac{\rme^{\rmi \vec k \cdot (\vec x - \vec x')}}
{ \frac{\hbar^2 \vec k^2}{2M} + \frac{1}{2} M \omega^2
\rho^2
- \mu + g n(\rho) + \rmi \gamma
}.
\label{eq:K1}
\end{multline}
Here $\gamma$ is an infinitesimal small positive number and $E_{c'}$
is a second energy cutoff that can be chosen different from $E_c$. For
large enough $E_{c'}$ the sum of first two terms in Eq.~(\ref{eq:K1})
is convergent. The inhomogeneous wavevector cut-off $k_{c'}(\rho)$ can
be found from
\begin{equation}
 \frac{\hbar^2 k^2_{c'}(\rho)}{2M}
+ \frac{1}{2}M\omega^2 \rho^2 - \mu + g n(\rho)
 = E_{c'}.
\end{equation}
The last term of Eq.~(\ref{eq:K1}) contains the singularity. One can
now perform the integration over $k$ analytically, which in the limit
$\vec x \rightarrow \vec x'$ gives \cite{Bulgac02, Grasso03}
\begin{equation}
  G_{\uparrow \downarrow}(\vec x, \tau; \vec x', \tau)
  = 
  -\frac{M \Delta(\vec x)}{4 \pi \hbar^2} \frac{1}{\vert \vec x - \vec x' \vert}
 + G_{\uparrow \downarrow}^{\mathrm{reg}}(\vec x, \tau; \vec x, \tau),
\end{equation}
where
\begin{multline}
G_{\uparrow \downarrow}^{\mathrm{reg}}(\vec x, \tau; \vec x, \tau)
 = \nu_s(\vec x) - \Delta(\vec x) K(\vec x, \vec x; E_{c'})
\\
\quad 
+ \frac{\Delta(\vec x)M}{2\pi^2 \hbar^2}
\left[
k_{c'}(\rho) -
\frac{1}{2}
k_F(\rho)
\log \left(
\frac{k_{c'}(\rho) + k_F(\rho)}
{k_{c'}(\rho) - k_F(\rho)}
\right) 
\right].
\label{eq:Greg}
\end{multline}
Here we have introduced the length of the Fermi wavevector $k_F(\rho)$
which can be found from the equation
\begin{equation}
\frac{\hbar^2 k^2_F(\rho)}{2M}
 =  
\mu - g n(\rho) - \frac{1}{2}M\omega^2 \rho^2 - \rmi \gamma.
\label{eq:KF}
\end{equation}
To obtain $\Delta(\vec x)$ we have to multiply Eq.~(\ref{eq:Greg}) by
$g$ as follows from Eq.~(\ref{eq:delta}).

Inserting Eq.~(\ref{eq:trlogGinv}) into Eqs.~(\ref{eq:thermopotB}) and
(\ref{eq:FHelmholtz}) gives the Helmholtz free energy
\begin{equation}
\begin{split}
F
&= 
- 
\sum_{i} \left[ \frac{ \vert E_i \vert}{2}   
+ \frac{1}{\beta} \log \left(1+\rme^{-\beta \vert E_i \vert} \right)
\right]
+ \sum_{i} \epsilon_i
\\
& \quad
- 
\int \rmd^3 x \,
G_{\uparrow \downarrow}(\vec x, \tau; \vec x, \tau)^*
\Delta(\vec x)
\\ 
& \quad
- 
g \int \rmd^3 x \, 
n_{\uparrow}(\vec x) n_{\downarrow}(\vec x )
+ \mu_\uparrow N_\uparrow + \mu_\downarrow N_\downarrow.
\label{eq:thermopotC1}
\end{split}
\end{equation}
Although some of the individual terms in the last equation are
ultraviolet divergent, their sum, and hence the Helmholtz free energy,
is ultraviolet finite.  The divergence present in the sum over the
eigenvalues of the Bogoliubov-de Gennes matrix is canceled by the sum
over the eigenvalues of the Hartree-Fock Hamiltonian and by the
logarithmic divergence originating from $G_{\uparrow
  \downarrow}$. This can be made clearer by expressing $G_{\uparrow
  \downarrow}$ in terms of $\nu_s$ and $K$. One finds 
\begin{equation}
\begin{split}
F
&
= 
-
\sum_{\vert E_i \vert < E_c} 
\left[ \frac{ \vert E_i \vert}{2}   
+ \frac{1}{\beta} \log \left(1+\rme^{-\beta \vert E_i \vert} \right)
\right]
+\sum_{\epsilon_i < E_c} \epsilon_i
\\
&\quad
+ \int \rmd^3 x \, K(\vec x, \vec x; E_c) \vert \Delta( \vec x)\vert^2
- 
\int \rmd^3 x \,
\nu_s(\vec x)^*
\Delta(\vec x)
\\
&
\quad - 
g \int \rmd^3 x \, 
n_{\uparrow}(\vec x) n_{\downarrow}(\vec x )
+ \mu_\uparrow N_\uparrow + \mu_\downarrow N_\downarrow.
\label{eq:FreeEnergyC}
\end{split}
\end{equation}
In the absence of a trapping potential $E_i$ is known analytically; then
it can be seen that Eq.~(\ref{eq:FreeEnergyC}) is ultraviolet
finite.  As we will see in the next section, $F$ is also finite in the
general case.

\section{Solving the Bogoliubov-de Gennes-equation}
\label{sec:sbdg}
In the previous section we have reduced the Dyson-Schwinger equation
to a nonlinear equation of the form 
\begin{equation}
(\Delta(\vec x), n_{\uparrow,
  \downarrow}(\vec x), \mathcal{N}_{\uparrow, \downarrow}) 
  = F(\Delta(\vec x), n_{\uparrow,
  \downarrow}(\vec x), \mu_{\uparrow, \downarrow}).
\label{eq:genDS}
\end{equation}
We will now discuss how to compute the function $F$ in practice. In
the next section we will solve the Dyson-Schwinger equation.

First we will use the symmetries of our problem to simplify the
analysis. For zero rotation frequency the superfluid is in a
vortex-free phase. Then, the pairing field $\Delta(\vec x)$ will be a
function of the radial coordinate $\rho$ only, i.e.\ $\Delta(\vec x) =
\tilde \Delta(\rho)$, where $\tilde \Delta(\rho) \in \mathbb{R}$.

We will assume that the first vortex that appears when increasing the
rotation frequency carries one unit of angular momentum and is located at the
center of the trap. The pairing field for such vortex has the following
form: $\Delta(\vec x) = \tilde \Delta(\rho) \exp(\rmi \phi)$. After the
single vortex has appeared, a vortex lattice can be formed by further
increasing the rotation frequency.

For these reasons we will make the following ansatz for the pairing
field
\begin{equation}
 \Delta(\vec x) = \tilde \Delta(\rho) \exp(\rmi k \phi),
\end{equation}
where $k$ is the winding number (unit of angular momentum) of the
vortex at the center. Hence the $k=0$ case corresponds to the
vortex-free phase. To determine the onset of the vortex phase, we have
to compare the Helmholtz free energies with $k = 0$ and $k =
1$. Because of the cylindrical symmetry of the trap and the fact that
a possible vortex is located at the origin, the number densities are a
function of $\rho$ only, i.e.  $n_{\uparrow, \downarrow}(\vec x) =
n_{\uparrow, \downarrow}(\rho)$.

In this case the solutions of the Bogoliubov-de Gennes equation,
Eq.~(\ref{eq:BdGB}), are of the following form
\begin{eqnarray}
 u_i(\vec x) &=&
\frac{1}{\sqrt{L}}
\rme^{\rmi p_z z}
\frac{1}{\sqrt{2\pi}}
\rme^{\rmi m \phi}
\frac{1}{\sqrt{\lambda \rho}}
u_{nmp_z}(\rho),
\label{eq:ui}
\\
 v_i(\vec x) &=&
\frac{1}{\sqrt{L}}
\rme^{\rmi p_z z}
\frac{1}{\sqrt{2\pi}}
\rme^{\rmi (m - k) \phi}
\frac{1}{\sqrt{\lambda \rho}}
v_{nmp_z}(\rho),
\label{eq:vi}
\end{eqnarray}
which can be verified by inserting these expressions into
Eq.~(\ref{eq:BdGB}). This also yields the Bogoliubov-de
Gennes equation for $u_{nmp_z}(\rho)$ and $v_{nmp_z}(\rho)$,
\begin{equation}
\mathcal{H}_{m p_z}
\left(
\begin{array}{c}
u_{nmp_z}(\rho)
\\
v_{nmp_z}(\rho)
\end{array}
\right)
= 
E_{nmp_z}
\left(
\begin{array}{c}
u_{nmp_z}(\rho)
\\
v_{nmp_z}(\rho)
\end{array}
\right),
\label{eq:BdGC}
\end{equation}
where
\begin{multline}
\mathcal{H}_{mp_z} = \\
\left(
\begin{array}{cc}
 H_m(\Omega) \!-\! \mu_\uparrow \! + \! g n_{\downarrow}(\rho)
&
\tilde \Delta(\rho)
\\
\tilde \Delta(\rho)
&
- H_{k-m}(\Omega) \!+\! \mu_\downarrow  \!-\! g n_{\uparrow}(\rho)
\end{array}
\right),
\end{multline}
and
\begin{equation}
\begin{split}
 H_m(\Omega) &
= \frac{\hbar^2}{2M} \left( -
  \frac{\mathrm{d}^2}{\mathrm{d} \rho^2}
    -\frac{1}{4\rho^2}
+ \frac{m^2}{\rho^2 + \lambda^2 \eta^2}
  \right)
\\
& \quad
+
\frac{1}{2} M \omega^2 \rho^2
+
\frac{p_z^2}{2M}
  - m \hbar \Omega.
\label{eq:HspB}
\end{split}
\end{equation}
Due to the factor $1/\sqrt{\rho}$ in Eqs.~(\ref{eq:ui}) and
(\ref{eq:vi}) $H_m(\Omega)$ is a bit different from the single
particle Hamiltonian defined in Eq.~(\ref{eq:Hsp}). We have inserted
this factor for later convenience. Furthermore, by inserting the term
$\lambda^2 \eta^2$ we have modified the centrifugal potential in such
a way that it becomes regular at $\rho = 0$. The original centrifugal
potential is reproduced for $\eta \rightarrow 0$. As we explain in
Appendix~\ref{app:barh}, this slight modification is necessary for
the numerical computation of the wavefunctions and energies. In our
computations we have taken $\eta = 10^{-5}$ and checked that the
results are completely stable if $\eta$ is varied by orders of
magnitude around this value.

From the normalization condition on $u_i(\vec x)$ and $v_i(\vec x)$ it
follows that $u_{nmp_z}(\rho)$ and $v_{nmp_z}(\rho)$ have to be
normalized as
\begin{equation}
\frac{1}{\lambda} 
\int_0^{\infty} \rmd \rho \,
 \left[
 u_{nmp_z}(\rho)^2
+
v_{nmp_z}(\rho)^2
\right] = 1.
\label{eq:bdgnorm}
\end{equation}

To solve the Bogoliubov-de Gennes equation numerically, we have to
discretrize Eq.~(\ref{eq:BdGC}). One can do this by expanding the
wavefunctions $u(\rho)$ and $v(\rho)$ in a certain basis. For
practical purposes, this basis has to be truncated, i.e.,  we will
represent $u_{nmp_z}(\rho)$ and $v_{nmp_z}(\rho)$ by a finite number $N$
of basis functions $\ell_i(\rho)$. More specifically we will write
\begin{equation}
 u_{nmpz}(\rho) = \sum_{i=1}^{N} c_i \ell_i(\rho) \,
\;\;\;\;
 v_{nmpz}(\rho) = \sum_{i=1}^{N} d_i \ell_i(\rho),
\end{equation}
here $c_i$ and $d_i$ are the expansion coefficients, which depend on
the quantum numbers $n$, $m$ and $p_z$. We will require the basis
functions to be orthonormal in the following way
\begin{equation}
 \frac{1}{\lambda} 
\int_0^{\infty} \rmd \rho \,
\ell_i(\rho)
\ell_j(\rho) = \delta_{ij}.
\label{eq:normell}
\end{equation}
From Eq.~(\ref{eq:bdgnorm}) it then follows that
the expansion coefficients have to be normalized as
\begin{equation}
 \sum_{i=1}^N \left(c_i^2 + d_i^2 \right) = 1.
\end{equation}

Equation~(\ref{eq:BdGC}) can now be transformed into an ordinary
eigenvalue equation for a $2N \times 2N$ matrix which reads
\begin{equation}
\left(
\begin{array}{cc}
\bar H_{m} - \bar \mu_\uparrow
&
\bar \Delta
\\
\bar \Delta
&
- \bar H_{k-m} + \bar \mu_\downarrow
\end{array}
\right)
\left(
\begin{array}{c}
 c
\\
d
\end{array}
\right)
= 
E_{nmp_z}
\left(
\begin{array}{c}
c
\\
d
\end{array}
\right).
\label{eq:BdGD}
\end{equation}
Here $\bar H$, $\bar \mu$ and $\bar \Delta$ are $N \times N$ real
symmetric matrices which are given by
\begin{eqnarray}
 (\bar H_{m})_{ij} &=& \frac{1}{\lambda}
\int_0^{\infty} \rmd \rho \,
\ell_i(\rho)
 H_m(\Omega) 
\ell_j (\rho),
\label{eq:hbar}
\\
 (\bar \mu_{\alpha})_{ij} &=& \frac{1}{\lambda}
\int_0^{\infty} \rmd \rho \,
\ell_i(\rho)
\left[
\mu_{\alpha}  - g n_{-\alpha}(\rho)
\right]
\ell_j (\rho), 
\label{eq:mubar}
\\
 (\bar \Delta)_{ij} &=& \frac{1}{\lambda}
\int_0^{\infty} \rmd \rho \,
\ell_i(\rho)
\tilde \Delta(\rho)
\ell_j (\rho).
\label{eq:deltabar}
\end{eqnarray}
Similarly we define $(\bar H_{\mathrm{HF}})_{ij}$ as the
representation of the Hartree-Fock Hamiltonian defined in
Eq.~(\ref{eq:HHF}).  Once these matrices have been computed
explicitly, Eq.~(\ref{eq:BdGD}) can be solved numerically using
standard linear algebra routines. From the solutions the wavefunctions
and the expressions for $\nu_s(\vec x)$ and $n_{\uparrow, \downarrow}(\vec x)$
can be constructed. We will give the explicit expressions at the end
of this section.

Since truncating the basis is an approximation, we have to choose the
basis carefully in order to make sure that the basis functions can
describe the exact solution to good accuracy. We can always improve
the accuracy of the truncation by taking a larger value of $N$, but the
drawback is that this increases the computational cost as well.

A good basis has to be able to describe the wave-functions in the case
$\Delta = 0$, which are the radial single particle wave functions
given in Eq.~(\ref{eq:rwhsp}) times $\sqrt{\rho}$. For that reason we
will choose a basis of the following form
\begin{equation}
 \ell_i(\rho) = \sqrt{w(\rho / \lambda)} \, l_i(\rho / \lambda),
\end{equation}
where $w(x) = x \exp(-x^2)$ and $l_i(x)$ is a set of linear
independent polynomials of degree $N-1$. In this basis all single
particle wave functions can be represented exactly by a finite number
of basis functions. For the calculation of vortices this is a
desirable feature, because in that case mixing between states with
different angular momentum quantum numbers occurs.

From Eq.~(\ref{eq:normell}) it follows that the polynomials $l_i(x)$
have to be chosen orthogonal with respect to weight function $w(x)$,
i.e.\
\begin{equation}
 \int_0^{\infty} \rmd x \, w(x) l_i(x) l_j(x) = \delta_{ij}.
\label{eq:lnorm}
\end{equation}

The set of polynomials of increasing degree that are  orthonormal to 
each other with the weight function $x^p \exp(-x^2)$ on the interval $[0,
\infty)$ are called Maxwell polynomials~\cite{Shizgal81}. We will
write the Maxwell polynomials for $p=1$ as $\phi_i(x)$, where $i = 0,
1, 2, \ldots$ denotes the degree.

One could have chosen as basis functions the Maxwell polynomials
directly, for example $l_i(x) = \phi_{i-1}(x)$. However, the
computation of the Bogoliubov-de Gennes matrix, Eq.~(\ref{eq:BdGD}),
becomes much easier if we use Lagrange interpolating functions as
basis functions. These Lagrange interpolating functions are a
particular linear combination of Maxwell polynomials and will be
specified below.  This approach is generally known as the Discrete
Variable Representation (DVR) method, or alternatively as the Lagrange
mesh discretization \cite{Light85, Baye86, Szalay93}. By applying this
method one can obtain highly accurate values for the energies and
wavefunctions~\cite{Baye02}.  The DVR method can be applied to any set
of orthogonal polynomials~\cite{Baye99}, and is not restricted to
Maxwell polynomials. To our knowledge, this is the first time that the
DVR method is used with Maxwell polynomials.

The DVR method is based on the Gaussian quadrature formula
\begin{equation}
 \int_0^\infty \rmd x\, 
w(x) f(x) \approx \sum_{n=1}^{N} w_n f(x_n).
\label{eq:gaussquadrature}
\end{equation}
Here and in the following the nodes $x_n$ are the roots of the Maxwell
polynomial $\phi_N(x)$ and $w_n$ the corresponding quadrature
weights. The integration formula Eq.~(\ref{eq:gaussquadrature}) is
exact for all polynomials $f(x)$ of degree less than $2N$.  From the
properties of the orthogonal polynomials, one can show that all $N$
roots are real and all weights are positive.  The nodes and the
weights are the only non-trivial properties of the Maxwell
polynomials one needs to know in order to apply the DVR method. Since
the Maxwell polynomials are non-standard polynomials, we review the
computation of its nodes and weights in Appendix~\ref{app:maxpoly}.

In the DVR method one chooses the functions $l_i(x)$ to be the
Lagrange interpolating functions through the nodes $x_n$.  Explicitly,
these functions read
\begin{equation}
 l_i(x) =  \frac{1}{\sqrt{w_i}} {\prod_{n=1, n \neq i}^{N}} 
\frac{x-x_n}{x_i - x_n}.
 \label{eq:lip}
\end{equation}
It follows directly that the polynomials $l_i(x)$ satisfy the
following useful property, $l_i(x_j) = \delta_{ij} /
\sqrt{w_i}$. Because the combination $l_i(x) l_j(x)$ is a
polynomial of degree $2N-2$, the Gaussian quadrature is exact for this
combination.  Hence we can use the Gaussian quadrature to show that
$l_i(x)$ satisfies the required orthonormality condition given in
Eq.~(\ref{eq:lnorm}),
\begin{equation}
 \int_0^{\infty} \rmd x \, w(x) l_i(x) l_j(x) 
= \sum_{k=1}^{N} \frac{w_k}{\sqrt{w_i w_j}}
\delta_{ik} \delta_{jk} 
= \delta_{ij}.
\end{equation}

We will now compute the matrices that appear in the Bogoliubov-de
Gennes equation, Eqs.~(\ref{eq:hbar})-(\ref{eq:deltabar})
explicitly. In the DVR method one uses the Gaussian quadrature to
compute the integrals. In this way we find that
\begin{eqnarray}
(\bar \mu_\alpha)_{ij} 
&\approx&
\mu_\alpha \delta_{ij} - g n_{-\alpha}(\lambda x_i) \delta_{ij},
\\
(\bar \Delta)_{ij}
&\approx& \tilde \Delta(\lambda x_i ) \delta_{ij}.
\end{eqnarray}
Here one of the nice features of the DVR method appears. All parts of
the Bogoliubov-de Gennes matrix that contain no derivatives become
diagonal and are easy and fast to evaluate. Furthermore, the values of
$\tilde \Delta(\rho)$ and $n_{\uparrow, \downarrow}(\rho)$ have 
to be evaluated only at the mesh points $\rho = \lambda x_i$. These mesh
points are unevenly spaced.

Another attractive feature of the DVR method is that matrix elements of
derivative operators can be computed exactly. We discuss the
computation of $\bar H$ in Appendix~\ref{app:barh}.  The exact
expression for the matrix $\bar H$ can be obtained by inserting
Eqs.~(\ref{eq:AijFinal}) and (\ref{eq:BijFinal}) into
Eq.~(\ref{eq:barHFinal}). While $\bar H$ becomes much more complicated
than $\bar \mu$ and $\bar \Delta$, this is not a disadvantage, because
we only need to compute $\bar H$ once.  This is in contrast to $\bar
\mu$ and $\bar \Delta$ which will change from iteration to iteration
when solving the Dyson-Schwinger equations.

In order to describe a single particle wave function with quantum
numbers $n$ and $m$ in this basis exactly, we need to take $N \geq 2n
+ \vert m \vert$.  For a given value of $N$ therefore only the lowest
$(N-\vert m \vert)/2$ eigenvalues and corresponding eigenvectors can
be computed exactly in this basis. 

Let us now order the eigenvalues of the Bogoliubov-de Gennes matrix
and the Hartree-Fock Hamiltonian, i.e.\ label them in such a way that
\begin{eqnarray}
 E_{nm p_z} &\leq& E_{n+1,mp_z},\,\,\,n=1 \ldots 2N,
\\
\epsilon_{nmp_z} &\leq& \epsilon_{n+1, mp_z},\,\,\,\,\, n=1 \ldots N.
\end{eqnarray}
If we can compute the first $n_{\mathrm{max}}$ eigenvalues of the
single particle Hamiltonian accurately then the eigenvalues of the
Bogoliubov-de Gennes matrix with $n_{\mathrm{i}} \leq n \leq
n_{\mathrm{f}}$ where $n_{\mathrm{i}} = N-n_{\mathrm{max}} + 1$ and
$n_{\mathrm{f}} = N + n_{\mathrm{max}}$, will be computed
accurately. We will write the maximum $n$ quantum number as
\begin{equation}
 n_{\mathrm{max}} = \left \lfloor 
\frac{\kappa N - \mathrm{max}
(\vert m \vert, \vert k -m \vert)
}{2} 
\right \rfloor,
\end{equation}
where $\lfloor x \rfloor$ denotes the floor function, and $\kappa$ is
a free parameter which equals $1$ if all single particle eigenvalues
and vectors need to be computed exactly.  Typically the next few
eigenvalues and eigenvectors can also be computed very reliably,
although they deteriorate rapidly at some point.  A larger value of
$\kappa$ is advantageous because more eigenvalues and vectors are
taken into account in the same basis. From our experience one can
safely take $1 \leq \kappa < 1.2$. The angular quantum number is
varying between $m_{\mathrm{min}}$ and $m_\mathrm{max}$. These are 
in the case of $k \geq 0$ given by
\begin{eqnarray}
 m_{\mathrm{min}} &=& -\lfloor \kappa N \rfloor + k, \\
 m_{\mathrm{max}} &=& \lfloor \kappa N \rfloor.
\end{eqnarray}

To solve the Bogoliubov-de Gennes equation we only need to evaluate
$\nu_s(\rho)$ and $n_{\uparrow, \downarrow}(\rho)$ at $\rho = \lambda
x_i$. At the mesh points the wave functions become rather simple
and read
\begin{eqnarray}
  u_{nmpz}(\lambda x_i) &=& c_i \sqrt{\frac{x_i}{w_i}} \exp(-x_i^2/2),
\\
  v_{nmp_z}(\lambda x_i) &=& d_i \sqrt{\frac{x_i}{w_i}} \exp(-x_i^2/2).
\end{eqnarray}
Combining now the last two equations with Eqs.~(\ref{eq:ui}) and
(\ref{eq:vi}) we find that at the mesh points the densities read
\begin{eqnarray}
  n_{\uparrow}(\lambda x_i) &=&
\frac{\rme^{-x_i^2}}{2\pi \lambda^2 w_i}
\sumint 
\sum_{n=n_{\mathrm{i}}}^{n_{\mathrm{f}}}
f(E_{nmp_z})
c_{i}^2,
\label{eq:densUpDVR}
\\
  n_{\downarrow}(\lambda x_i) &=&
\frac{\rme^{- x_i^2}}{2\pi \lambda^2 w_i}
\sumint
\sum_{n=n_{\mathrm{i}}}^{n_{\mathrm{f}}}
f(-E_{nmp_z})
d_{i}^2.
\label{eq:densDownDVR}
\end{eqnarray}
Here we introduced the symbol
\begin{equation}
\sumint \equiv
\sum_{m=m_{\mathrm{min}}}^{m_{\mathrm{max}}}
\int_0^{p_c} 
\frac{\mathrm{d}p_z}{\pi \hbar},
\end{equation}
where $p_c$ is a cutoff on the $p_z$ integration. Furthermore, we used the 
fact that all integrands are symmetric in $p_z$. We have performed
the integration over $p_z$ numerically using the adaptive Simpson
method. After that we performed the sum over $m$.

The function $\nu_s(\vec x)$ from which $\Delta(\vec x)$ can be
obtained becomes at the mesh points 
\begin{multline}
 \nu_s(\lambda x_i) =
\rme^{\rmi k \phi}
\frac{ \rme^{-x_i^2} }
{2\pi\lambda^2 w_i}
\sumint 
\left[
\sum_{n=n_{\mathrm{i}}}^{n_{\mathrm{f}}}
f(E_{nmp_z}) c_{i} d_{i}
\right.
\\
+
\left.
 \tilde 
\Delta(\lambda x_i)
\sum_{n=1}^{n_\mathrm{max}} \frac{\tilde c_i^2}{2\epsilon_{nmp_z}} 
\theta(\epsilon_{nmp_z} - E_s)
\right]
,
\end{multline}
where $\epsilon_{nmp_z}$ and $\tilde c_i$ are respectively the
eigenvalues and eigenvectors of the matrix representation of the
Hartree-Fock Hamiltonian, $(\bar H_{\mathrm{HF}})_{ij}$.  To compute
$\Delta(\lambda x_i)$ we also need to evaluate $K(\lambda x_i, \lambda
x_i, E_{c'})$, which is defined in Eq.~(\ref{eq:Kx}). To obtain $K$ only
the eigenvalues and eigenvectors of the Hartree-Fock Hamiltonian in
the case $p_z = 0$ have to be computed numerically. Their values for
nonzero $p_z$ follow trivially, since $\epsilon_{nmp_z} =
\epsilon_{nm0} + p_z^2/(2M)$. We then obtain
\begin{multline}
 K(\lambda x_i, \lambda x_i; E_{c'}) =
\frac{\rme^{-x_i^2}}{2\pi \lambda^2 w_i}
\sum_{m=m_{\mathrm{min}}}^{m_{\mathrm{max}}} 
\sum_{n=1}^{n_\mathrm{max}} \tilde c_{i}(p_z=0)^2
\\
\times
\int \frac{\mathrm{d} p_z}{2\pi \hbar}
\left. 
\frac{1}{2 \epsilon_{nmp_z}} \right
\vert_{E_s < \epsilon_{nmp_z} < E_{c'}}
.
\label{eq:KDVR}
\end{multline}
The value of $E_{c'}$ is limited by the largest $\epsilon_{nmp_z}$ one
can compute reliably. An estimate of this value is $E_{c'} < \kappa N
- \mu + gn(0)$.  The integration over $p_z$ in Eq.~(\ref{eq:KDVR}) can
straightforwardly be performed analytically (we will however not write
down the result here). Therefore $K(\lambda x_i, \lambda x_i; E_{c'})$
can be computed much faster than $\nu_s(\lambda x_i)$ for which
numerical integration over $p_z$ is required. To compute $K(\lambda
x_i, \lambda x_i; E_{c'})$ we can thus afford to use a much larger
basis (we have taken $N=128$) than for $\nu_s(\vec x)$. A larger basis
implies that we can choose a larger value of the cutoff $E_{c'}$
leading to more reliable answers. The values of $K$ on a coarser mesh
can then be obtained from interpolation.

From Eq.~(\ref{eq:Greg}) we obtain the pairing field
\begin{eqnarray}
\Delta(\lambda x_i)
 &=& g \nu_s(\lambda x_i) - g \Delta(\lambda x_i) 
  K(\lambda x_i, \lambda x_i; E_{c'})
\nonumber \\
&&
+ g \frac{\Delta(\lambda x_i)M}{2\pi^2 \hbar^2}
\Biggl[
k_{c'}(\lambda x_i)
\label{eq:DeltaDVR}
\\
&& - \frac{1}{2}
k_F(\lambda x_i)
\log \left(
\frac{k_{c'}(\lambda x_i) + k_F(\lambda x_i)}
{k_{c'}(\lambda x_i) - k_F(\lambda x_i)}
\right) 
\Biggr].
\nonumber 
\end{eqnarray}

The Helmholtz free energy per unit of harmonic oscillator length in
the $z$-direction follows from Eq.~(\ref{eq:FreeEnergyC}) and reads
\begin{equation}
\begin{split}
\mathcal{F}
&
= 
- \lambda
\sumint \Biggl \{
\sum_{n=n_{\mathrm{i}}}^{n_{\mathrm{f}}}
\Bigl[ \frac{ \vert E_{nmpz} \vert}{2}  
+ \frac{1}{\beta} \log \left(1+\rme^{-\beta \vert E_{nmpz} \vert} \right)
\Bigr]
\\
& \quad \quad
-
\sum_{n=1}^{n_{\mathrm{max}}} \Bigl[
\epsilon_{nmpz}
+ \frac{ \theta(\epsilon_{nmpz}\!-\!E_s) }{2 \epsilon_{nmpz}} 
\sum_{i=1}^{N}
\tilde c_i^2 \vert \Delta(\lambda x_i)\vert^2
\Bigr]
\Biggr \}
\\
& \quad
- 
2\pi \lambda^3 \sum_{i=1}^{N} w_i
\nu_s(\lambda x_i)^* \Delta(\lambda x_i)
\rme^{x_i^2}
\\
&
\quad - 
g 2 \pi \lambda^3
\sum_{i=1}^{N} w_i
n_{\uparrow}(\lambda x_i ) n_{\downarrow}(\lambda x_i)
\rme^{x_i^2}
+ \mu_\uparrow \mathcal{N}_\uparrow + \mu_\downarrow \mathcal{N}_\downarrow.
\label{eq:FreeEnergyDVR}
\end{split}
\end{equation}
Here we have used the Gauss-Maxwell quadrature to compute the
integrals over $\rho$. It can be shown numerically that the integrand
of $\mathcal{F}$ decreases rapidly for large $p_z$, making
$\mathcal{F}$ ultraviolet finite.

Another important quantity which now can be constructed is the
expectation value of the angular momentum operator $L_z$. With help of
the number densities, Eqs.~(\ref{eq:densUpDVR}) and
(\ref{eq:densDownDVR}), the angular momentum density can be
written as 
\begin{multline}
l_z(\lambda x_i) =
\frac{\rme^{-x_i^2} \hbar}{2\pi \lambda^2 w_i}
\sumint 
\sum_{n=n_{\mathrm{i}}}^{n_{\mathrm{f}}}
\left[
m c_{i}^2 f(E_{nmp_z})
\right.
\\
\left.
+ (k - m) d_{i}^2  f(-E_{nmp_z})
\right] .
\end{multline}
Integrating the last equation over $\rho$ and $\phi$ gives the angular
momentum per unit harmonic oscillator length in the $z$-direction
which we will denote by $\mathcal{L}_z$.

\section{Solving the Dyson-Schwinger equation}
\label{sec:ds}
Solving the Dyson-Schwinger equation amounts to finding the solution
of Eqs.~(\ref{eq:densUpDVR}), (\ref{eq:densDownDVR}) and
(\ref{eq:DeltaDVR}), together with the constraint on the number of
particles. Schematically the equation to be solved is of the form
given in Eq.~(\ref{eq:genDS}).  Such equation can be solved using a
multidimensional root-finding method. For that purpose we will use the
Newton-Broyden method \cite{Broyden65}, which leads to very fast
convergence once close to the solution. As an input to the
Newton-Broyden method one should provide initial guesses for $\Delta$,
$n_{\uparrow, \downarrow}$, and $\mu_{\uparrow, \downarrow}$ and also
provide the Jacobian of $F$.

To obtain the initial conditions we will use the Thomas-Fermi
approximation.  The Thomas-Fermi approximation to the density 
is found by solving the following equation
\begin{multline}
 n_{\uparrow, \downarrow}(\rho) 
= 
 \int 
\frac{\mathrm{d}^3 p}{
(2\pi \hbar)^3}
f
\bigl
(\frac{p^2}{2M} + \frac{1}{2} M \omega^2 \rho^2 - \Omega L_z 
\\
- \mu_{\uparrow, \downarrow} + g n_{\downarrow, \uparrow}(\rho)
\bigr).
\end{multline}
If $T=0$ the last equation becomes
\begin{equation}
  n_{\uparrow, \downarrow}(\rho) = 
\frac{1}{6 \pi^2 \lambda^3}
\left[
2 \frac{\mu_{\uparrow, \downarrow} - g n_{\downarrow, \uparrow}(\rho)}
{\hbar \omega}
-
\left(1 - \frac{\Omega^2}{\omega^2} \right) \frac{\rho^2}{\lambda^2}
\right]^{3/2}.
\label{eq:TFdens}
\end{equation}
This equation can easily be solved numerically. If $\mu_{\uparrow} =
\mu_{\downarrow}$ the Thomas-Fermi radius (the minimal $\rho$ for
which $n_{\uparrow, \downarrow}(\rho) = 0$) of the gas becomes
\begin{equation}
 R = \frac{1}{1-\Omega^2/\omega^2} \sqrt{\frac{2 \mu}{\hbar \omega}}
\lambda.
\label{eq:TFradius}
\end{equation}
The initial values for $\mu_{\uparrow, \downarrow}$ can be obtained by
integrating the Thomas-Fermi density profiles and solving numerically
for the desired $\mathcal{N}_{\uparrow, \downarrow}$.

For the initial condition to the pairing field in the case $k=0$ we use
the result of the BCS theory in the weak coupling limit (see
e.g.~Ref.~\cite{Grasso03}),
\begin{equation}
 \tilde \Delta(\rho) = 4 k_F(\rho)^2 \lambda^2
\exp \left(-2 - \frac{\pi}{2 k_F(\rho) \vert a \vert} \right)
\hbar \omega,
\label{eq:BCSpairing}
\end{equation}
where $k_F(\rho)$ is defined in Eq.~(\ref{eq:KF}). As an initial
condition for nonzero $k$, we multiply this equation by a factor
$1-\exp[-\rho/(k\xi)]$ where the BCS coherence length equals
\begin{equation}
\xi = \frac{k_F(0) \lambda^2}
{\pi \tilde \Delta(0) / (\hbar \omega)}.
\end{equation}

In the Newton-Broyden method the Jacobian has to be computed at each
iteration. We have computed the initial Jacobian using finite
differences. If $\mu_\uparrow = \mu_\downarrow$ there are $2(2N+1)$
evaluations of $F$ required, so obtaining the initial Jacobian is
computationally expensive.

However, it is not necessary to apply finite differences to recompute
the Jacobian in the next iteration. The next Jacobian can be obtained
from the previous one using a Broyden update \cite{Broyden65}. Such
update has negligible computational cost. After the initial Jacobian
has been acquired Newton-Broyden iterations typically reach
convergence in about ten iterations. The time it takes to reach
convergence is completely dominated by the time it takes to compute
the initial Jacobian. The computational cost of the whole problem
scales roughly with $N^5$, a factor of $N^3$ originates from solving
the Bogoliubov-de Gennes equation, a factor of about $N$ arising from
the sum over angular quantum numbers and another factor of $N$
from the Jacobian.

Given the solution, the Helmholtz free energy can straightforwardly
computed by applying Eq.~(\ref{eq:FreeEnergyDVR}).

\section{Results}
\label{sec:results}
In the numerical computations the following parameters were used $\eta
= 10^{-5}$, $\kappa = 1.1$, $E_s = 2.43 \hbar \omega$, $p_c / \hbar =
20 / \lambda$ and $E_{c'} = 100 \hbar \omega$. The adaptive Simpson
integration was performed with a relative precision goal of $10^{-5}$
and absolute precision goal $10^{-10}$ for each individual component
of $\Delta(\lambda x_i)$ and $n_{\uparrow, \downarrow}(\lambda
x_i)$. The accuracy goals for the computation of the initial Jacobian
were set much lower, which speeds up the computation. As long as
these goals are not set too low they will not ruin the convergence of
the Newton-Broyden algorithm. We would like to stress that a less
accurate Jacobian that can make the Newton-Broyden algorithm converge
does not influence the accuracy of the final values of $\Delta(\lambda
x_i)$ and $n_{\uparrow, \downarrow}(\lambda x_i)$.

The iterations of the Newton-Broyden algorithm were performed until
the relative difference between the norm of $(\tilde \Delta(\lambda
x_i), n_{\uparrow,\downarrow}(\lambda x_i), \mathcal{N}_{\uparrow,
  \downarrow} )$ and $F(\Delta(\lambda x_i), n_{\uparrow,
  \downarrow}(\lambda x_i), \mu_{\uparrow, \downarrow})$ became less
than $10^{-7}$. Frequently, a relative accuracy of $10^{-9}$ could be
reached.

The number of basis functions $N$ was chosen such that convergence was
reached. At least all energy levels below the Fermi energy have to be
computed accurately. Since a larger number of particles implies a
larger Fermi energy, for a larger number of particles a larger value
of $N$ is required.  We have done computations from $N=16$ to
$N=80$. A calculation with $N=80$ took several days on a single modern
CPU. For calculations with $\mathcal{N} = 100$, $200$ and $1000$ we
typically used respectively $N=40$, $48$ and $64$.

We have checked that our results are completely stable under
acceptable variation of these parameters. By studying these
variations, we have convinced ourselves that the largest values of
$\Delta(\lambda x_i)$, $n_{\uparrow, \downarrow}(\lambda x_i)$ have a
relative accuracy of a least $10^{-3}$. The free energy could be
obtained with a relative accuracy of $10^{-5}$.

\begin{figure}[t]
 \includegraphics{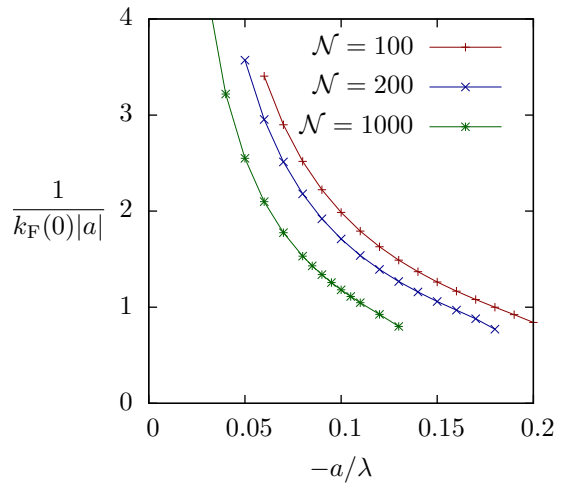}
 \caption{Inverse interaction strength at the center of the trap as a
   function of scattering length, for $\Omega = 0$ and different
   number densities of particles $\mathcal{N}$.}
\label{fig:interactionstrength}
\end{figure}

Furthermore we have taken $T=0$ throughout and considered the
situation that the number of particles per unit harmonic oscillator
length in the $z$ direction in each species is equal,
i.e.~$\mathcal{N}_\uparrow = \mathcal{N}_\downarrow =
\mathcal{N}/2$. We have investigated situations with $\mathcal{N} =
100$, $200$ and $1000$, different scattering lengths and different
rotation frequencies. The scattering lengths we have considered
correspond to inverse interaction strengths at the center of the trap
in the range $1 \lesssim 1/(k_{\mathrm{F}}(0) \vert a \vert) \lesssim
4$. We have depicted their relationship at zero rotation frequency in
Fig.~\ref{fig:interactionstrength}. In this range the
Hartree-Fock-Bogoliubov approximation is expected to be valid. If one
wants to study stronger interactions one has to take into account the
higher order diagrams in order to get a reliable result.

To get an idea of the scales in a typical experiment, we can use that
in Ref.~\cite{Zwierlein05} a Fermi gas made out of $^6\mathrm{Li}$
atoms was studied in a trapping potential with radial frequency
$\omega/(2 \pi) = 57\;\mathrm{Hz}$. This situation corresponds to
$\lambda \sim 5.4\; \mu\mathrm{m}$ and $\hbar \omega / k_B \sim
2.7\;\mathrm{nK}$.

We will now first give a detailed overview of the results at zero
rotation frequency. After that we will discuss the effects of rotating
the trap and present the main objective of this work, the critical
rotation frequency for vortex formation.

\subsection{Zero rotation frequency}
\begin{figure}[t]
 \includegraphics{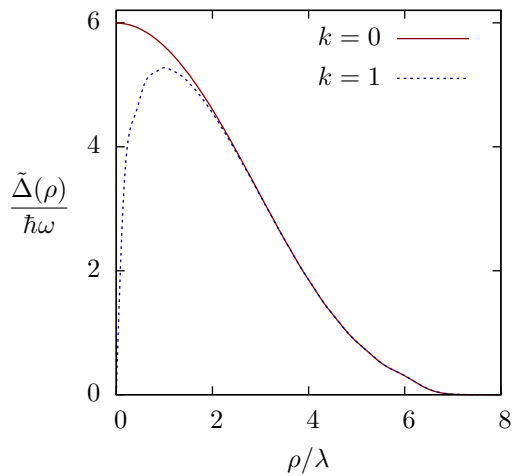} 
 \caption{Pairing field as a function of radius, for
   $a=-0.1\lambda$, $\mathcal{N}=1000$ and $\Omega = 0$.}
 \label{fig:d1000}
\end{figure}

\begin{figure}[t]
 \includegraphics{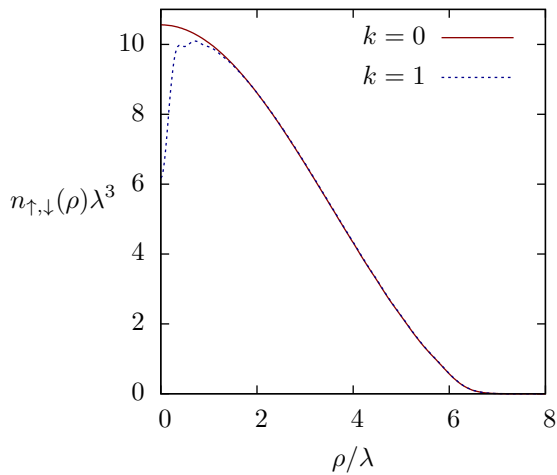}
 \caption{Number density as a function of radius, for $a =
   -0.1\lambda$, $\mathcal{N} = 1000$ and $\Omega = 0$.  }
 \label{fig:n1000}
\end{figure}
In Fig.~\ref{fig:d1000} we compare the pairing field of the
vortex-free superfluid ($k=0$) with the pairing field of the vortex
with unit angular momentum ($k=1$).  These pairing fields were
computed for $\mathcal{N}=1000$ and $a = -0.1 \lambda$ with
$N=64$. Because the data points lie so close to each other, we have
only displayed a line that interpolates through the data points for
visibility reasons. Since the rotation frequency was taken to be zero,
the vortex is metastable.

The pairing field clearly vanishes at the center of the vortex. Away
from the vortex core the pairing field is restored to its value in the
$k=0$ case. The typical distance at which this happens is
the BCS coherence length $\xi$. This implies that the size of the vortex
core grows when decreasing the strength of the interaction. For very
weak interactions, $\xi$ can become larger than the radius of the
gas. In that case even a metastable vortex is no longer possible.  In
the second part of this section we will study the size of the vortex
core at the critical rotation frequency for vortex formation in some
detail.

The vortex also leaves its imprint on the corresponding density
profiles which are displayed in Fig.~\ref{fig:n1000}.  As already
found in Refs.~\cite{Bulgac03, Feder04}, the density at the center of
the trap is significantly depleted in the presence of a vortex. To
compensate for the removal of particles at the center, the gas will
expand. This is a tiny effect and is, therefore, not visible in
Fig.~\ref{fig:n1000}. In the next subsection we will study density
depletion at the critical rotation frequency.

The density profile for normal pairing is
very well described by the Thomas-Fermi approximation, which is the
solution of Eq.~(\ref{eq:TFdens}). The pairing field however only
agrees qualitatively with the Tomas-Fermi approximation,
Eq.~(\ref{eq:BCSpairing}), as was also concluded in
Ref.~\cite{Bruun99}.

\begin{figure}[t]
 \includegraphics{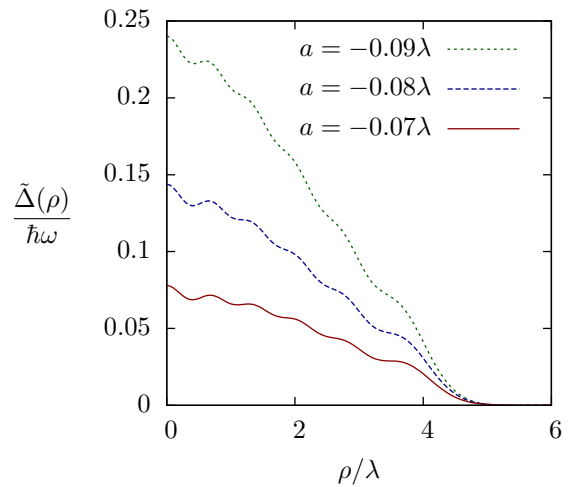} 
 \caption{Pairing field as a function of radius, for
   $\mathcal{N}=100$, $k=0$ and different scattering lengths.}
\label{fig:d100}
\end{figure}

We have displayed pairing field profiles for $\mathcal{N} = 100$ in
Fig.~\ref{fig:d100}. The interaction strength was taken to be weak, in
the range $0.07 \le \vert a\vert/\lambda\le 0.09$, which leads to
small pairing fields. In such situations we encountered 
oscillations in the pairing field. To ensure that this is not a
numerical artifact, we have compared these profiles computed with
$N=32$, $N=48$ and $N=64$. We find that they are completely consistent
with each other. The oscillations in the pairing field are only
prominent in the case of a small number of particles with weak
interactions.
\begin{figure}
 \includegraphics{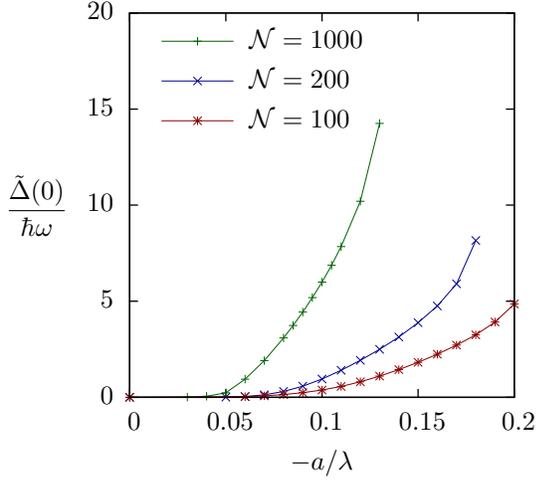} 
\caption{Pairing field at the center of the trap,
as function of the scattering length, for
different number of particles and $\Omega = 0$.}
\label{fig:centralPairing}
\end{figure}

Let us now study the effect of variation of the scattering length and
the number of particles in the zero rotation limit. In 
Fig.~\ref{fig:centralPairing} we display the pairing field at the
center of the trap as a function of the scattering length. Clearly,
increasing the interaction strength and the number of particles
(i.e. the Fermi wave number) both lead to larger pairing fields.  This is
qualitatively in agreement with the BCS pairing formula,
Eq.~(\ref{eq:BCSpairing}).

\begin{figure}
 \includegraphics{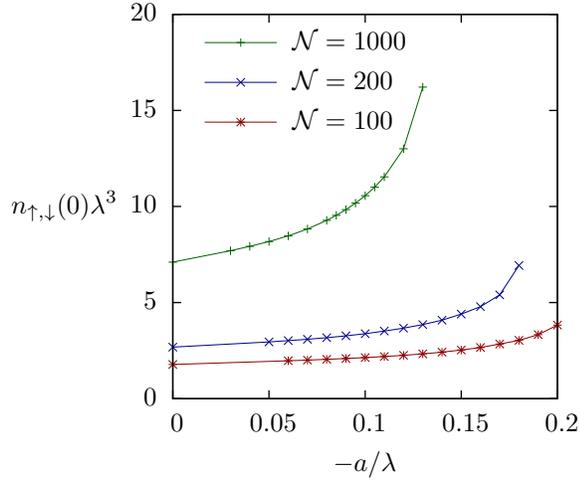}
 \caption{
Density at the center of the trap,
as function of the scattering length, for
different number of particles, and $\Omega = 0$.
}
\label{fig:centralDensity}
\end{figure}

In Fig.~\ref{fig:centralDensity} we have displayed the corresponding
number density at the center of the trap. Increasing the number of
particles leads naturally to a larger number density at the center.
Stronger attractive interactions lead to a more compressed gas, which
likewise results in a larger density at the center.

\begin{figure}
 \includegraphics{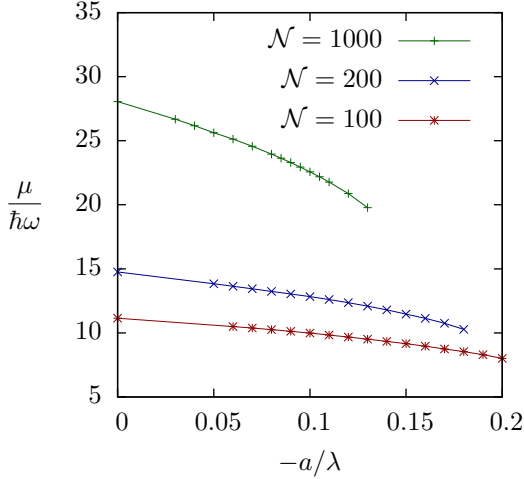} 
 \caption{ Chemical potential as function of the scattering length,
   for different number of particles, and $\Omega = 0$.}
\label{fig:mu}
\end{figure}

In Fig.~\ref{fig:mu} we have displayed the corresponding chemical
potential. Obviously a larger number of particles implies a larger
chemical potential. The chemical potential decreases with
increasing the strength of the interaction. This is because the
Hartree term $-g n_{\downarrow, \uparrow}(\rho)$, which acts as a sort
of inhomogeneous chemical potential, grows with increasing the
interaction strength.

The radius of the gas can, to very good approximation, be obtained by
inserting the values of the chemical potential in the Thomas-Fermi
estimate, Eq.~(\ref{eq:TFradius}). This gives at zero rotation
frequency $R/\lambda = \sqrt{2 \mu / \hbar \omega}$. It then follows
from Fig.~\ref{fig:mu} that increasing the number of particles
increases the radius. On the other hand, increasing the interaction
strength reduces the radius. The gas becomes more compressed because
the interaction between the two components is attractive.

\subsection{Non-zero rotation frequency}

\begin{figure}
 \includegraphics{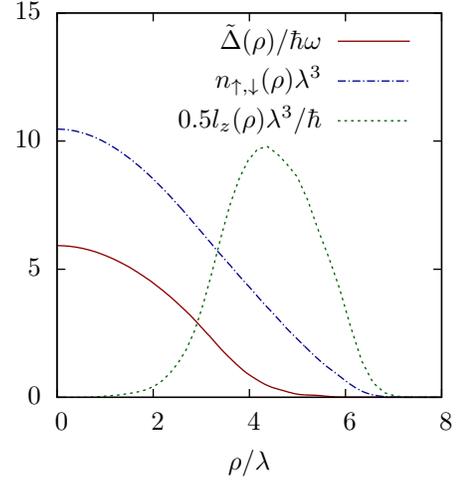} 
 \caption{Pairing field, number density, and angular momentum density
   as a function of radius. The results correspond to a vortex-free
   superfluid with $\mathcal{N} = 1000$, $a = -0.1 \lambda$, and
   $\Omega = 0.15 \omega$.}
\label{fig:rot_k0}
\end{figure}

\begin{figure}
 \includegraphics{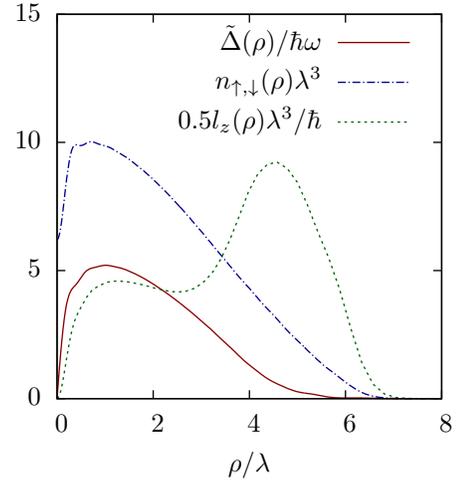} 
 \caption{Same as in Fig.~\ref{fig:rot_k0}, but for a vortex with $k=1$.
}
\label{fig:rot_k1}
\end{figure}

To illustrate the effects of rotation, we have displayed the pairing
field, the number density and the angular momentum density for $\Omega
= 0.15 \omega$, $\mathcal{N} = 1000$, and $a=-0.1\lambda$ in
Fig.~\ref{fig:rot_k0} ($k=0$) and Fig.~\ref{fig:rot_k1} ($k=1$).
These figures can be compared to the results at zero rotation
frequency which are displayed in Figs.~\ref{fig:d1000} and
\ref{fig:n1000}.  As we will show below, $\Omega=0.15 \omega$ is the
critical rotation frequency for vortex formation in this situation.

A vortex-free superfluid state cannot carry angular momentum. For that
reason the angular momentum density vanishes in the region where the
pairing field is sizable, as can be seen in Fig.~\ref{fig:rot_k0}.
Since angular momentum density appears at large $\rho$, it indicates
the presence of unpaired fermions at the edges of the gas. Another way
to observe this effect is that for large radial coordinates, the
pairing field disappears before the number density does. It can be
seen in Fig.~\ref{fig:rot_k1} that a vortex generates angular momentum
density in the superfluid region. For the same reasons as in the $k=0$
case, unpaired fermions are present at the boundaries of the gas.

By careful comparison of Figs.~\ref{fig:rot_k0} and~\ref{fig:rot_k1}
one can observe that the pairing field of the vortex is slightly
larger in the outer region. This is generally the case and leads in
addition to the effects mentioned in the introduction to a fourth
contribution to the energy difference between a vortex and a
vortex-free phase.  Rotation increases the radius of the cloud as
well. However, at this rotation rate this is only a very small effect
and is therefore not visible in the figures.

Now let us discuss the determination of the critical rotation
frequencies for unpairing and vortex formation. To obtain these
frequencies we have computed the Helmholtz free energy. The phase with
the lowest free energy is the preferred phase.

\begin{figure}
 \includegraphics{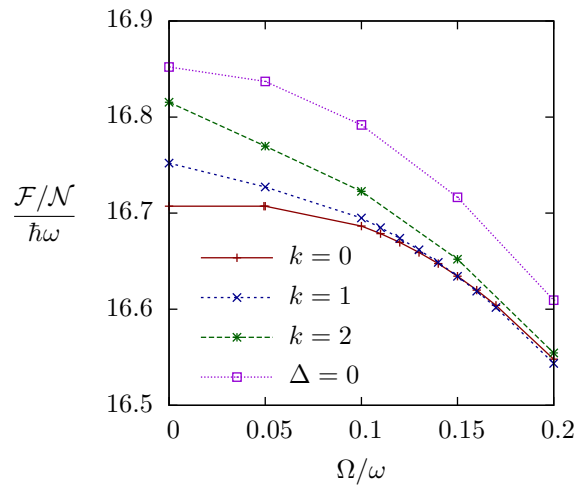} 
 \caption{Helmholtz free energy divided by the number of particles,
  as a function of rotation frequency,
   for $a = -0.1 \lambda$ and $\mathcal{N} = 1000$.  The label $k=0$
   corresponds to a superfluid without vortices, the nonzero values of
   $k$ correspond to a single vortex at the center of the trap with
   angular momentum $k$.  The label $\Delta = 0$ corresponds to the
   situation in which all fermions are unpaired.}
\label{fig:F1000}
\end{figure}

\begin{figure}
 \includegraphics{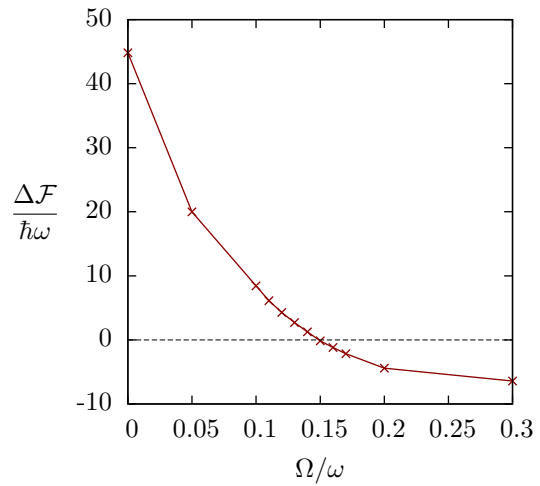} 
 \caption{Difference in Helmholtz free energy per unit length
  in the $z$-direction
  between the vortex phase
   with $k=1$ and the vortex-free phase with $k=0$, as a function of
   rotation frequency, for $a = 0.1 \lambda$ and $\mathcal{N} =
   1000$.}
\label{fig:DF1000}
\end{figure}

In Fig.~\ref{fig:F1000} we have displayed the Helmholtz free energy
divided by the number of particles, for $\mathcal{N} = 1000$ and $a =
-0.1 \lambda$. A number of interesting features of the superfluid are
shown in this figure.  First of all, the superfluid phase is always
preferred over the unpaired phase, since $\Delta = 0$ has the largest
free energy. Furthermore, for $\Omega < 0.05 \omega$ the gas forms a
vortex-free superfluid. It can be seen that in this region the free
energy does not depend on the rotation frequency. This indicates that
the entire gas is in a superfluid state. However, for $\Omega > 0.05
\omega$ the free energy of the vortex-free phase starts to decrease
when increasing the rotation frequency. This implies that the gas has
acquired angular momentum, which occurs via unpairing the fermions at
the edges of the gas. Hence for $0.05 \omega < \Omega < 0.15 \omega$
the gas forms a vortex-free superfluid with unpaired fermions at the
boundaries. At $\Omega > 0.15 \omega$ a superfluid with a $k=1$ vortex
becomes the preferred phase. One can see this more clearly in
Fig.~\ref{fig:DF1000}, where we have displayed the difference in free
energy between the vortex phase with $k=1$ and the vortex-free
phase. The critical rotation frequency can be found from interpolation
of the data points which in this case yields $\Omega_c = 0.149
\omega$. For $\Omega < 0.15 \omega$ the $k=1$ phase is metastable.
Because superfluids with a vortex carry angular momentum, the
derivative of their free energy with respect to rotation frequency is
negative, even at zero rotation frequency. In the unpaired phase this
derivative vanishes at zero frequency, because the fully unpaired gas
does not contain angular momentum at zero frequency.

At zero temperature, one can also compute the critical rotation
frequency for unpairing in a more direct way.  At zero rotation
frequency all quasi-particle excitations (except the superfluid
phonon) are gapped, i.e.~$\vert E_{nmp_z} \vert > 0$. As follows from
the discussion in Appendix~\ref{app:BdG}, if $\mu_\uparrow =
\mu_\downarrow$ and $\Omega = 0$, both $E_{nmp_z}$ and $-E_{nmp_z}$
are eigenvalues of the Bogolibuov-de Gennes matrix. Rotation shifts
these eigenvalues downwards by $m \hbar \Omega$. As long as no gapless
mode arises the rotational contributions from positive and negative
energies cancel so that this shift has no effect on the free
energy. For that reason the free energy for $k=0$ stays constant up to
a certain rotation frequency. Only when the first gapless mode
appears, the free energy will change. The minimal rotation frequency
at which this occurs is the critical rotation frequency for unpairing,
$\Omega_u$. Thus this rotation frequency can be found from the
solutions at $\Omega = 0$ in the following way
\begin{equation}
 \Omega_{u} = \mathrm{min}
\left \vert
\frac{1}{m\hbar} E_{nmp_z}(\Omega = 0)
\right \vert,
\label{eq:omegau}
\end{equation}
where the minimum is to be taken over all values of $n$, $m$, and
$p_z$. Determination of $\Omega_u$ in this way is computationally much
less expensive than obtaining it from the free energy.

\begin{figure}
 \includegraphics{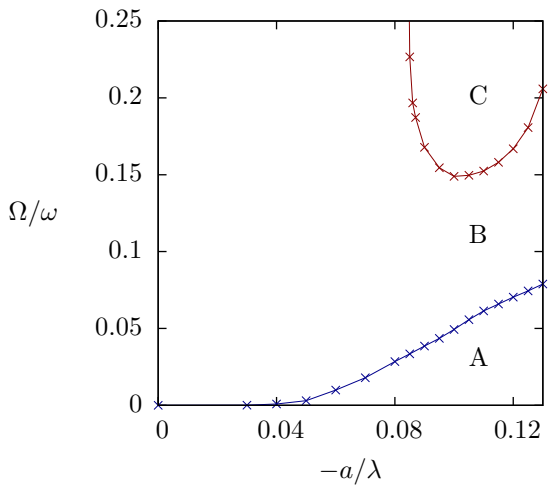} 
 \caption{Phase diagram of a two-component Fermi gas as a function of
   scattering length and rotation frequency, for $\mathcal{N} =
   1000$ and $T=0$. The lines correspond to the phase boundaries.  The label A
   indicates that the entire gas is in a vortex-free superfluid state.
   The label B indicates a vortex-free superfluid in the center with
   unpaired fermions in the outer regions of the gas.  The label C
   indicates a superfluid with vortices in the center and unpaired
   fermions in the outer regions.}
\label{fig:PD1000}
\end{figure}

\begin{figure}
 \includegraphics{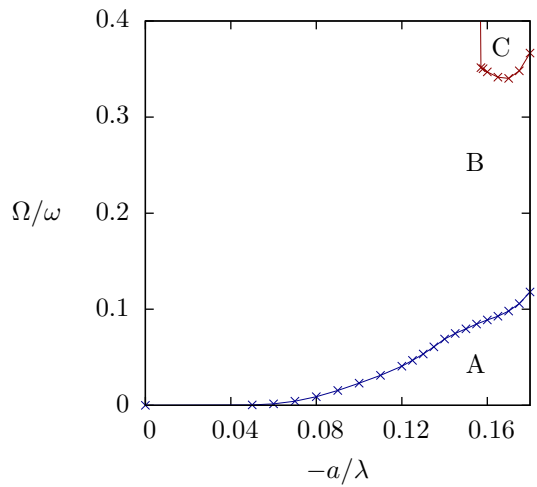} 
 \caption{Same as in Fig.~\ref{fig:PD1000}, but for $\mathcal{N} = 200$.  }
\label{fig:PD200}
\end{figure}

Let us now discuss our main result: the phase diagram as a function of
scattering length and rotation frequency.  In Figs.~\ref{fig:PD1000}
and \ref{fig:PD200} we have displayed these diagrams for
$\mathcal{N}=1000$ and $\mathcal{N}=200$ respectively.

There are two transitions in these phase diagrams.  The lower line
denotes the unpairing transition. The order parameter corresponding to
this transition is the angular momentum. This transition is of second
order since the angular momentum changes continuously.  At $T \neq 0$
this transition turns into a crossover. Hence at $\Omega=\Omega_u$ and
$T=0$ the gas resides at a so-called quantum critical point. Above
this critical point the order parameter behaves as $\mathcal{L}_z \sim
t^{\beta}$ where $t=(\Omega - \Omega_u)/\Omega_u$. We find numerically
that the critical exponent has the value $\beta = 1$.

The upper line denotes the critical rotation frequency for the
formation of a vortex with unit angular momentum at the center of the
trap. Energy arguments suggest that with increasing rotation frequency
the first vortex configuration that will nucleate is a single vortex
with $k=1$. A single vortex with $k > 1$ will have lager energy, as
can be seen in Fig.~\ref{fig:F1000}.  Several vortices with $k=1$ have
again larger energy and their nucleation would require larger than
critical rotation frequencies.  Therefore, the upper line shows the
critical rotation frequency for vortex formation. The order parameter
corresponding to this transition is the winding number of the
vortex. Since this winding number changes discontinuously, this
transition is of first order.

Increasing the absolute value of the scattering length leads to a
larger critical rotation frequency for unpairing.  Furthermore, for a
given scattering length $\Omega_u$ becomes larger when the number of
particles is increased. Both effects can be explained by the fact that
a stronger bound pair is more difficult to break.

As can be seen from the phase diagrams, we find that vortices are
formed only for relatively large negative scattering lengths. The
critical rotation frequency for vortex formation has a minimum at a
certain intermediate value of the scattering length. This minimum
arises from the interplay of two effects. Firstly, the energy cost of
creating a vortex at zero rotation frequency increases with increasing
the negative scattering length. This explains the rise of the critical
frequency at large negative scattering lengths.  The second effect is
caused by the difference in energy gain due to rotation. The $k=1$
phase will always have a larger rotational energy gain than the $k=0$
phase due to the angular momentum generated by the vortex. However,
for small interaction strengths above the unpairing transition, the
difference between these gains is relatively small. This is because in
this case it is relatively easy to break the pairs at the boundaries
of the gas, which contribute to the rotational energy gain in both the
$k=0$ and $k=1$ phase. As a result of this effect the critical
frequency increases for small negative scattering lengths. For a
certain small scattering length the difference in rotational energy
gain cannot overcome the costs associated to the vortex.  For this
reason at small negative scattering lengths the vortex phase has an
abrupt transition to a vortex-free phase. We see that below $a \approx
-0.155 \lambda$ for $\mathcal{N}=200$ and $a \approx -0.085 \lambda$
for $\mathcal{N}=1000$ vortex formation does not occur at all the
rotation frequencies displayed in the phase diagram.

Vortex formation sets in at a lower rotation rate when the number
density of particles is increased from $\mathcal{N}=200$ to
$\mathcal{N}=1000$. For the number of particles we have investigated
we find that the vortices always appear together with unpaired
fermions at the edges of the gas. One could speculate that for a
larger number of particles $\Omega_c$ will be reduced so that a vortex
phase will appear before unpairing at the edges could become possible.
In other words, we anticipate that for a large number of particles the
phase diagram might feature a direct phase transition from the A to
the C phase without the intermediate B phase.

\begin{figure}
 \includegraphics{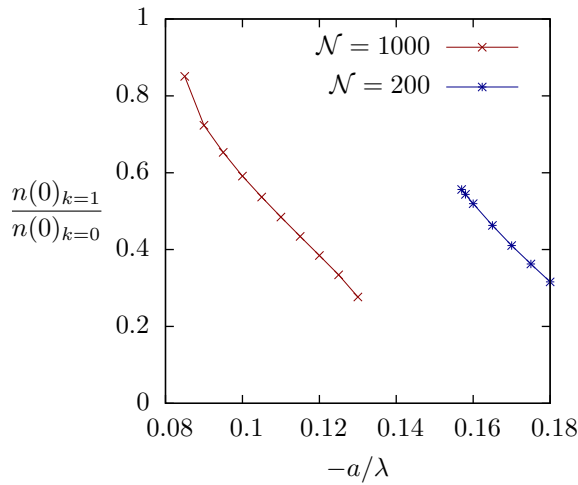} 
 \caption{Relative density depletion at the core of a vortex, as a
   function of scattering length, at the critical rotation frequency
   for vortex formation.}
\label{fig:dd}
\end{figure}

In Fig.~\ref{fig:dd} we have displayed the central number density of a
vortex over the central density of a vortex-free superfluid, at the
transition to vortex formation. It can be seen that the amount of
density depletion is relatively small for weak interactions and grows
with increasing the interaction strength.

\begin{figure}
 \includegraphics{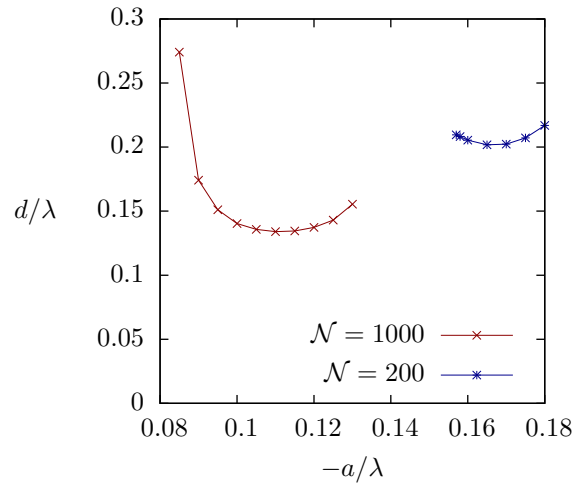} 
 \caption{Half-width of the vortex as a function of scattering length,
   at the critical rotation frequency for vortex formation. }
\label{fig:vs}
\end{figure}

Let us define the half-width $d$ of the vortex to be the radius at
which $\tilde \Delta(\rho)_{k=1} / \tilde \Delta(\rho)_{k=0} = 1/2$.
In Fig.~\ref{fig:vs} we have displayed this half-width at the
transition for vortex-formation. For $\mathcal{N}=1000$ it can be
clearly seen that weak interactions lead to larger vortices, which is
caused by the increase of the BCS coherence length.

\section{Conclusions}
\label{sec:conclusions}

In this article we studied a two-component Fermi gas with attractive
$s$-wave interactions confined in a cylindrically symmetric harmonic
trap. Our key results are summarized in a phase diagram spanned by the
rotation frequency and scattering length for zero temperature and a
fixed number of particles.  Explicit results are shown for a number
density of 1000 and 200 particles per unit harmonic oscillator length
in the $z$-direction in Figs.~\ref{fig:PD1000} and \ref{fig:PD200}
respectively.

To obtain the phase diagram we have used the two-particle irreducible
effective action. We only took into account the leading order
diagrams, which is equivalent to the Hartree-Fock-Bogoliubov
approximation.  This constrains our study to interaction strengths of
magnitude $1/(k_F(0) \vert a \vert) \lesssim 1$.  The equations we
obtained were solved numerically using the DVR method based on Maxwell
polynomials. 

In the phase diagram three phases can be distinguished.  For small
rotation frequencies the entire gas forms a superfluid. At a certain
critical frequency a second order transition occurs to a superfluid
phase, which features unpaired fermions that are concentrated at the
edges of the gas. At this critical rotation frequency the gas resides
at a quantum critical point when the temperature vanishes.  For even
larger rotation frequencies vortices are formed via a first order
transition. These vortices only appear for large negative scattering
lengths.  We have found that at a certain scattering length the
critical rotation frequency for vortex formation has a minimum.

The presence of unpaired fermions at the boundaries of the gas results
in an increase of the critical rotation frequency for vortex
formation. For this reason one cannot use the free energy difference
between a vortex phase and vortex-free phase at zero rotation
frequency to compute the critical rotation frequency for vortex
formation.

Our theoretical findings can be compared to the experiment at will,
since the vortices have been observed in rotating two-component Fermi
gases~\cite{Zwierlein05, Zwierlein06}. It would be interesting to
obtain the structure of the phase diagram from such experiments.

The theoretical understanding of the phase diagram can be improved in
several ways.  It would be worthwhile to investigate how the phase
diagram is modified by temperature and by an imbalance in the number
of fermions. Furthermore, it would be very useful to extend our
analysis to larger interaction strengths, in order to obtain reliable
results in the unitary regime.

Finally we would like to point out that vortices can also be induced
by synthetic magnetic fields, as has been shown experimentally in a
Bose-Einstein condensate~\cite{Lin09}. Since such synthetic magnetic
field is similar to rotation, another interesting extension of our
work would be to compute the critical synthetic magnetic field
strength for vortex formation in a two-component Fermi gas.

\section*{Acknowledgments}
The work of H.J.W.\ was supported by the Alexander von Humboldt
Foundation. A.S. thanks the Deutsche Forschungsgemeinschaft for
partial support. We would like to thank Steven Knoop and Dirk
H. Rischke for useful discussions.

\appendix

\section{The 2PI effective action}
\label{app:2pi}
Consider an action $S = -\hbar \Psi^\dagger G_0^{-1} \Psi + S_{\mathrm{int}}$,
where $G_0^{-1}$ denotes the bare inverse propagator and
$S_{\mathrm{int}}$ is the part of the action that contains
interactions.  The partition function $Z[K]$ corresponding to this
action in the presence of a source term $K$ is given by
\begin{equation}
 Z[K] = \int \mathcal{D} \Psi^\dagger
\mathcal{D} \Psi
\exp \left[- S / \hbar + \
\Psi^\dagger K \Psi \right].
\end{equation}
Let us denote the exact propagator in the presence of a source
term $K$ as $G$.  The 2PI effective action is defined as
\cite{2PI}
\begin{equation}
 \Gamma[G] = - \log Z[K] + \mathrm{Tr}(K G),
 \label{eq:2pidef}
\end{equation}
where $K$ has to be chosen in such a way that exact propagator in the
presence of $K$ equals $G$. Since the exact propagator has to satisfy
the Dyson-Schwinger equation, we can conclude that $K = G^{-1} -
G_0^{-1} + \Sigma[G]$, where $\Sigma[G]$ is the 1PI self-energy. Taking
the derivative of Eq.~(\ref{eq:2pidef}) with respect to $G$ gives
$\frac{\delta \Gamma[G]}{\delta G} = K$, so that in the extremal
points $\Gamma[G] = -\log Z[0]$.  Inserting the expression for $K$ in
Eq.~(\ref{eq:2pidef}) gives
\begin{equation}
 \Gamma[G] = - \log Z_2 - 
\mathrm{Tr} \left( G_0^{-1} G - 1 \right)
+ \mathrm{Tr}(\Sigma[G] G),
\end{equation}
where $Z_2 = Z[G^{-1} - G_0^{-1} + \Sigma[G]]$, which implies that
$Z_2$ is the partition function of a theory with action $S_2 = - \hbar 
\Psi^\dagger G^{-1} \Psi + S_{2,\mathrm{int}}$, where $S_{2,\mathrm{int}} =
- \hbar \Psi^\dagger \Sigma[G] \Psi + S_{\mathrm{int}}$. One can now compute
$-\log Z_2$ in a perturbative series, using $G$ as the propagator.
It follows that the first two terms in this perturbative series are
\begin{equation}
 - \log Z_2 = - \mathrm{Tr} \log G^{-1} - 
 \mathrm{Tr} \left( \Sigma[G]G \right)
+ \ldots.
\end{equation}
Because the 1PI self-energy is now included in the interaction term,
cancellations will occur such that only the 2PI diagrams survive
\cite{2PI}. Hence
\begin{equation}
 \Gamma[G] = - \mathrm{Tr} \log G^{-1} - 
\mathrm{Tr} \left( G_0^{-1} G - 1 \right)
+ \Gamma_2[G],
\end{equation}
where $\Gamma_2[G]$ is now the sum of all 2PI diagrams generated
by the interaction $S_{\mathrm{int}}$ with propagator $G$.

\section{Derivation of the Bogoliubov-de Gennes-equation}
\label{app:BdG}
The inverse of a non-singular Hermitian matrix $A$ can be obtained
from its eigenvalues $\lambda_n$ and the corresponding orthonormal
eigenvectors $\vert n \rangle$, with $\langle n \vert m
\rangle= \delta_{nm}$. Putting the $n$-th eigenvector in the
$n$-th column of a new matrix $U$, one finds that $U$ is unitary,
i.e. $U^\dagger U = 1$ and $A = U \Lambda U^\dagger$, where $\Lambda =
\mathrm{diag}(\lambda_1, \lambda_2, \ldots)$.  The inverse of $A$ can
now be constructed as $A^{-1} = U \Lambda^{-1}
U^\dagger$, where $\Lambda^{-1} = \mathrm{diag}(1/\lambda_1,
1/\lambda_2, \ldots)$. By performing the matrix multiplications, the
last equation can be conveniently written as $A^{-1} = \sum_{n} \vert
n \rangle \langle n \vert / \lambda_n$. A single component of the
inverse matrix now reads $A^{-1}_{ij} = \sum_{n}
\langle i \vert n \rangle \langle n \vert j \rangle/ \lambda_n$.

The inverse Nambu-Gor'kov propagator $G^{-1}$, Eq.~(\ref{eq:dsB}), can
be written as $-\hbar G^{-1} = \hbar \partial / \partial \tau +
\mathcal{H}$, where the Hermitian matrix $\mathcal{H}$ is given
by Eq.~(\ref{eq:hbdg}).

Since $\mathcal{H}$ is independent of $\tau$, the eigenfunctions of
$G^{-1}$ are a product of eigenfunctions of $\hbar \partial / \partial
\tau$ and eigenfunctions of $\mathcal{H}$. Because $G(X, X')$ has to
satisfy anti-periodic boundary conditions in imaginary time, the
properly normalized eigenfunctions of $\hbar \partial / \partial \tau$
are plane waves $\phi_m(\tau) = \exp(-\rmi \omega_m \tau) /
\sqrt{\hbar \beta}$ with eigenvalue $-\rmi \hbar \omega_m$, where the
Matsubara frequency $\omega_m = (2m+1) \pi / (\hbar \beta)$, with $m
\in \mathbb{Z}$. Let us denote the normalized eigenfunctions of
$\mathcal{H}$ as $ \left( u_i(\vec x), v_i(\vec x) \right)^T$ with
corresponding eigenvalue $E_i$ (which is real). This eigenvalue
equation which is given explicitly in Eq.~(\ref{eq:BdGB}) is known as 
the Bogoliubov-de Gennes equation \cite{Gennes64}. Normalization
($U^\dagger U = 1$) implies that
\begin{equation}
\int \rmd^3 x\, \left[ \vert
  u_i(\vec x) \vert^2 + \vert v_i(\vec x) \vert^2 \right] = 1.
\end{equation}
Furthermore from completeness ($U U^\dagger = 1$) one finds 
\begin{multline}
\sum_i
\left(
\begin{array}{cc}
u_i(\vec x) u_i^*(\vec x')
&
u_i(\vec x) v_i^*(\vec x')
\\
v_i(\vec x) u_i^*(\vec x')
&
v_i(\vec x) v_i^*(\vec x')
\end{array}
\right)
= \\
\delta(\vec x - \vec x')
\left(
\begin{array}{cc}
1
&
0
\\
0
&
1
\end{array}
\right).
\label{eq:completeness}
\end{multline}

We can now invert the inverse propagator $G^{-1}$, to obtain the
Nambu-Gor'kov propagator,
\begin{multline}
G(X, X')
=
-\frac{1}{\beta} 
\sum_i 
\sum_{m \in \mathbb{Z}}
\frac{1}{-\rmi \hbar \omega_m + E_i}
\rme^{-\rmi \omega_m (\tau - \tau')}
\\
\times
\left(
\begin{array}{cc}
u_i(\vec x) u_i^*(\vec x')
&
u_i(\vec x) v_i^*(\vec x')
\\
v_i(\vec x) u_i^*(\vec x')
&
v_i(\vec x) v_i^*(\vec x')
\end{array}
\right ).
\label{eq:FullNGprop}
\end{multline}
We can see from this equation explicitly that $G_{\downarrow
  \uparrow}(X, X') = G_{\uparrow \downarrow}(X', X)^*$. In order to
obtain the pairing field and the number densities we need to
evaluate $G(X, X'_\pm)$. Here $X'_\pm = (\vec x, \tau \pm \eta)$ with
$\eta$ an infinitesimal small positive number. In this limit one can
compute the sum over Matsubara frequencies exactly. After using the
completeness relation, Eq.~(\ref{eq:completeness}), one then finds
\begin{eqnarray}
G(X, X'_\pm)
&=&
\sum_{i}
f(E_i)
\left(
\begin{array}{cc}
u_i(\vec x) u_i^*(\vec x')
&
u_i(\vec x) v_i^*(\vec x')
\\
v_i(\vec x) u_i^*(\vec x')
&
v_i(\vec x) v_i^*(\vec x')
\end{array}
\right )
\nonumber \\
&& 
 - \theta(\mp) \delta(\vec x - \vec x')
\left(
\begin{array}{cc}
1
&
0
\\
0
&
1
\end{array}
\right),
\label{eq:G1}
\end{eqnarray}
where $f(E) = [\exp(\beta E) + 1]^{-1}$ denotes the Fermi-Dirac
distribution function and $\theta(x)$ is the unit-step function.  The
term proportional to the step function reflects the anti-commutation
relation of the fermionic operators. The sum over $i$ runs over all
eigenvalues. Using Eqs.~(\ref{eq:delta}), (\ref{eq:densup}), and
(\ref{eq:densdown}) one can now read off the expressions
for the pairing field and the number densities.

If $\mu_{\uparrow} = \mu_{\downarrow}$, then the densities of the two
species are equal so that $n_\uparrow(\vec x) = n_\downarrow(\vec
x)$. By taking the complex conjugate of Eq.~(\ref{eq:BdGB}) it follows
in this case that if $E_i$ is an eigenvalue of $\mathcal{H}$ with
eigenvector $ \left( u_i(\vec x), v_i(\vec x) \right)^T$, then also
$-E_i$ is an eigenvalue of $\mathcal{H}$ with eigenvector $\left(
  v^*_i(\vec x), -u^*_i(\vec x) \right)^T$. One can now use this fact
to restrict the sum over $n$ to eigenvectors with positive eigenvalues
only, so that if $\mu_{\uparrow} = \mu_{\downarrow}$ one has
\begin{multline}
G(X, X'_\pm)
=
\sum_{E_i \geq 0} 
f(E_i)
\left(
\begin{array}{cc}
u_i(\vec x) u_i^*(\vec x')
&
u_i(\vec x) v_i^*(\vec x')
\\
v_i(\vec x) u_i^*(\vec x')
&
v_i(\vec x) v_i^*(\vec x')
\end{array}
\right )
\\
+ \sum_{E_i \geq 0}
[1-f(E_i)]
\left(
\begin{array}{cc}
v^*_i(\vec x) v_i(\vec x')
&
-v_i^*(\vec x) u_i(\vec x')
\\
- u^*_i(\vec x) v_i(\vec x')
&
u_i^*(\vec x) u_i(\vec x')
\end{array}
\right )
\\
- \theta(\mp) \delta(\vec x - \vec x')
\left(
\begin{array}{cc}
1
&
0
\\
0
&
1
\end{array}
\right).
\label{eq:G2}
\end{multline}
From this equation one can read off the expressions for the number
density and pairing field as they often appear in the literature.
In this paper however we will solely use Eq.~(\ref{eq:G1}), because it
leads to more compact expressions and has broader validity. The only
slight disadvantage is that in Eq.~(\ref{eq:G1}) we have to sum over
all eigenvalues.

To evaluate the grand potential, Eq.~(\ref{eq:thermopotB}),
we need to compute $\mathrm{Tr} \log G^{-1}$.  Using that the trace of
a logarithm is the sum over the logarithm of the eigenvalues one 
finds
\begin{equation}
\frac{1}{\beta} \mathrm{Tr} \log
G^{-1}  = 
\frac{1}{\beta}  \sum_{i} \sum_{m \in \mathbb{Z}} 
\log( -\rmi \hbar \omega_m  + E_i).
\end{equation} 
In order to perform the sum over the Matsubara frequencies one adds
and subtracts the following infinite constant to the last equation
\begin{equation}
C = \frac{1}{\beta} \sum_{i} \left[ \sum_{m \in \mathbb{Z}} 
\log( -\rmi \hbar \omega_m) - \log(2) \right].
\end{equation}
After summing over Matsubara frequencies one finds
\begin{equation}
\frac{1}{\beta} \mathrm{Tr} \log
G^{-1} =
\sum_{i} \left[ \frac{ \vert E_i \vert}{2}   
+ \frac{1}{\beta} \log \left(1+\rme^{-\beta \vert E_i \vert} \right)
\right] + C.
\label{eq:trlogGinv}
\end{equation}
Since $C$ is independent of $E_i$ it shifts the thermodynamic
potential by an irrelevant constant and can therefore be ignored.  Now
the result Eq.~(\ref{eq:trlogGinv}) is not entirely correct.  For
example it is still infinite and in the limit of $\Delta(\vec x) = 0$
the grand potential of an unpaired Fermi gas is not obtained.  
To cure this problem one needs to take carefully the limit
$\eta \rightarrow 0$. We proceed as in Refs.~\cite{Stoof, Negele88} 
to obtain 
\begin{eqnarray}
\frac{1}{\beta} \mathrm{Tr} \log
G^{-1} 
&=&
\sum_{i} \left[ \frac{ \vert E_i \vert}{2}   
+ \frac{1}{\beta} \log \left(1+\rme^{-\beta \vert E_i \vert} \right)
\right]
\nonumber
\\
&&
- \sum_i \epsilon_i,
\label{eq:trlogGinvF}
\end{eqnarray}
here $\epsilon_i$ are the eigenvalues of the Hartree-Fock Hamiltonian,
which is defined in Eq.~(\ref{eq:HHF}). The last equation is indeed
finite and reduces to the grand potential of an unpaired Fermi gas in
the case $\Delta(\vec x) = 0$.

\section{Computation of nodes and weights of Maxwell polynomials}
\label{app:maxpoly}
Any set of orthonormal polynomials of increasing degree $i$ and hence 
also the Maxwell polynomials $\phi_i(x)$ satisfies the following 
recursion relation (see e.g.~Refs.~\cite{Szego39, Gautschi04}),
\begin{equation}
  \sqrt{\beta_{i+1}} \phi_{i+1}(x) =
(x - \alpha_{i}) \phi_{i}(x) - \sqrt{\beta_i} \phi_{i-1}(x).
\end{equation}
Here $\alpha_i$ and $\beta_i$ are the recursion coefficients. 

Once the recursion coefficients are known, the nodes (which are the
roots of $\phi_N(x)$) and weights of order $N$ can be found by solving
the following eigenvalue equation numerically (see
e.g.~Refs.~\cite{Szego39, Gautschi04}),
\begin{multline}
\left(
\begin{array}{cccc}
\alpha_0 & \sqrt{\beta_1} &   & \\
\sqrt{\beta_1} & \alpha_1 & \sqrt{\beta_2} &   \\
 & \ddots  & \ddots & \ddots \\
 &  & \sqrt{\beta_{N-1}} & \alpha_{N-1} \\
\end{array}
\right)
\left(
\begin{array}{c}
\phi_0(x_n) \\
\phi_1(x_n) \\
\vdots \\
\phi_{N-1}(x_n) \\
\end{array}
\right)
\\
=
x_n
\left(
\begin{array}{c}
\phi_0(x_n) \\
\phi_1(x_n) \\
\vdots \\
\phi_{N-1}(x_n) \\
\end{array}
\right).
\end{multline}
The eigenvalues of the last equation are the $N$ nodes $x_n$.  The
weights follow directly from the eigenvectors in the following way (see
e.g.~Refs.~\cite{Szego39, Gautschi04}),
\begin{equation}
  w_n = \left[\sum_{i=0}^{N-1} \phi_i(x_n)^2 \right]^{-1}.
\end{equation}
The eigenvectors have to be normalized in such a way that that the
orthonormality condition is satisfied, so that $\phi_0(x_n) =
[\int_0^{\infty} \rmd x \, w(x)]^{-1/2}$.

The recursion coefficients can be computed using the Stieltjes
procedure, see e.g.\ Ref.~\cite{Gautschi82}. Although the recursion
coefficients for the Maxwell polynomials can be computed analytically
in this way, this is impractical since the coefficients quickly become
extremely complicated. Hence we have computed the recursion
coefficients numerically. When doing so, one encounters another
problem. The Stieltjes algorithm is extremely-ill
conditioned~\cite{Gautschi82}, implying that small errors blow up
quickly. To avoid this, we followed Ref.~\cite{Shizgal81}, and
performed the Stieltjes algorithm using arbitrary precision
arithmetic. This can be done using for example the computer program
Mathematica. In order to compute all recursion coefficients up to
$N=128$ with $22$ digits accuracy (so that it fits in double
precision) we had to use 10,000 digits precision in the Stieltjes
procedure.

In this way we have computed the weights and nodes of the
Gauss-Maxwell quadrature with $p=1$ up to $N=128$. For $N=2, 4, 8$ and
$16$ we could compare with tables presented in Ref.~\cite{Shizgal81}.
We find excellent agreement, almost up to machine precision accuracy.

\section{Computation of $\bar H$}
\label{app:barh}
Here we will compute $\bar H$, which is defined in
Eq.~(\ref{eq:hbar}). The integration over the parts of $H_m(\Omega)$
that are independent of $\rho$ is straightforward so that we
can write
\begin{equation}
\begin{split}
 \left(\bar H \right)_{ij}
& = 
\frac{1}{2}
\hbar \omega
\left[-
(A)_{ij}
+ 
m^2 (B)_{ij}
\right]
\\
& \quad
- \hbar \Omega m \delta_{ij}
+ \frac{p_z^2}{2M} \delta_{ij},
\label{eq:barHFinal}
\end{split}
\end{equation}
where the matrices $A$ and $B$ read
\begin{eqnarray}
(A)_{ij} 
&=&
\left(
\lambda^2 
\frac{\rmd^2}{\rmd \rho^2}
+ \lambda^2 
\frac{1}{4 \rho^2}
- \frac{\rho^2}{\lambda^2}
\right)_{ij},
\\
(B)_{ij} 
&=&
\left( 
\frac{\lambda^2}
{\rho^2 + \eta^2 \lambda^2}
\right)_{ij}.
\end{eqnarray}
First we will compute the matrix $(A)_{ij}$. We find
\begin{equation}
\begin{split}
(A)_{ij}
 & = 
\int_0^\infty \rmd x \,
w^{1/2}(x) l_i(x) \\
& \quad \quad \quad \times
\left[
\frac{\rmd^2}{\rmd x^2}
+\frac{1}{4 x^2}
- x^2
\right]
w^{1/2}(x) l_j(x) 
\\
&
= \int_0^\infty \rmd x \,
w(x) l_i(x)
\\
& \quad \quad \quad \times
\left[
\frac{\rmd^2}{\rmd x^2}
+
\left(\frac{1}{x} - 2 x \right)
\frac{\rmd}{\rmd x}
- 2
\right]
l_j(x).
\label{eq:Aij1}
\end{split}
\end{equation}
To obtain the last line we have used that $w(x) = x \exp(-x^2)$.
Eq. (\ref{eq:Aij1}) can be rewritten as
\begin{eqnarray}
(A)_{ij} & =&
\int_0^\infty \rmd x \,
w(x) l_i(x)
\left[
\frac{\rmd^2}{\rmd x^2}
- 2 x \frac{\rmd}{\rmd x}
- 2
\right]
l_j(x)
\nonumber \\
&& 
+ 
\int_0^\infty \rmd x \,
w(x) l_i(x)
\frac{1}{x}
\left[
\frac{\rmd}{\rmd x}
l_j(x)
- l_j'(0) \right]
\nonumber \\
&& 
+ \,
l_j'(0) 
\int_0^\infty \rmd x \, w(x) \frac{1}{x} \left[l_i(x) - l_i(0) \right]
\nonumber \\
&& 
+ \,
l_i(0) l_j'(0) \int_0^{\infty} \rmd x\, \exp(-x^2). 
\label{eq:Aij2}
\end{eqnarray}
This form has the advantage that the integrands of the first three
terms are products of the weight function $w(x)$ and a
polynomial of degree less than $2N$. Hence we can evaluate these
terms exactly using the Gauss-Maxwell quadrature.  The last term of
Eq.~(\ref{eq:Aij2}) can also be computed analytically and is equal to
$\l_i(0) l_j'(0) \sqrt{\pi} / 2$.

To proceed we will first evaluate the first and second order derivatives
of $l_j(x)$ at the nodes $x_i$ which become (see also Ref.~\cite{Szalay93})
\begin{equation}
\begin{split}
\left.
\frac{\rmd }{\rmd x} l_j(x) 
\right \vert_{x=x_i}
&= \frac{1}{\sqrt{w_i}}
\xi_i \delta_{ij}   
\\
& \quad
+
\frac{1}{\sqrt{w_i}}
\frac{C_i}{C_j} 
\frac{1}{x_i - x_j}(1 - \delta_{ij}),
\end{split}
\end{equation}
\begin{multline}
\left.
\frac{\rmd^2}{\rmd x^2} l_j(x) 
\right \vert_{x=x_i}
 =
\frac{1}{\sqrt{w_i}}
\left(\xi_i^2 - \zeta_i \right) \delta_{ij}
\\
+ \frac{1}{\sqrt{w_i}} \frac{C_i}{C_j}
\left[
\frac{ 2 \xi_i}{x_i - x_j} - \frac{2}{(x_i - x_j)^2}
\right]
( 1 - \delta_{ij}),
\end{multline}
where 
\begin{eqnarray}
\xi_i
&=&
\sum_{n=1, n \neq i}^{N} \frac{1}{x_i - x_n}, 
\\
\zeta_i
&=&
\sum_{n=1, n \neq i}^{N} \frac{1}{(x_i - x_n)^2},
\\
C_i &=& \sqrt{w_i} {\prod_{n=1,n\neq i}^N (x_i - x_n)}.
\end{eqnarray}
Furthermore $l_i(0)$ and $l_j'(0)$ are given by
\begin{eqnarray}
 l_i(0) &=& \frac{1}{\sqrt{w_i}} \prod_{n=1, n\neq i}^N 
\frac{x_n}{x_n - x_i}, 
\\
  l_j'(0) & =& - D_j l_j(0),
\\
D_j &=&
\sum_{n=1, n \neq j}^{N} \frac{1}{x_n}.
\end{eqnarray}
Putting everything together we find that
\begin{equation}
\begin{split}
(A)_{ij} & = \left[\xi_i^2
- \zeta_i
+ 
\left(\frac{1}{x_i}- 2 x_i \right) \xi_i
 - 2 \right]\delta_{ij} 
\\
& \quad
+ 
\frac{C_i}{C_j}
\left[
\frac{ 2 \xi_i}{x_i - x_j} - \frac{2}{(x_i - x_j)^2}
\right.
\\
&
\quad \quad \quad \quad 
+
\left.
\left(
\frac{1}{x_i} - 2 x_i
\right) \frac{1}{x_i - x_j}
\right](1-\delta_{ij})
\\
& \quad
+ D_j \left[
\sum_{n=1}^{N} \frac{w_n}{x_n}
- \frac{1}{2} 
\sqrt{\pi}
\right] l_i(0) l_j(0).
\label{eq:AijFinal}
\end{split}
\end{equation}
The matrix $A$ is symmetric as follows from Eq.~(\ref{eq:Aij1}),
although this is not directly clear from the last equation.

Now let us compute the matrix $(B)_{ij}$. Expressed
in terms of an integral over the basis functions this matrix
reads
\begin{equation}
(B)_{ij}
= \int_0^\infty \rmd x \,
w(x) l_i(x)
l_j(x) \frac{1}{x^2 + \eta^2}.
\end{equation}
We are interested in $(B)_{ij}$ in the limit of small $\eta$.
Therefore in the rest of the calculations, we will drop terms that are
of order $\eta$ and higher. Like in the calculation for $(A)_{ij}$ we
rewrite the integrand in such a way that we get terms which are a
polynomial of degree less than $2N$ times $w(x)$. We can then easily
integrate these terms using the Gauss-Maxwell quadrature. We can rewrite
$(B)_{ij}$ in the limit $\eta \rightarrow 0$ as
\begin{equation}
\begin{split}
(B)_{ij}
& = \int_0^\infty \rmd x \,
w(x) \left[l_i(x) - l_i(0) \right]
\left[l_j(x) - l_j(0) \right]
\frac{1}{x^2}
\\
&
\quad
+ l_i(0)
\int_0^\infty \rmd x \,
w(x) \left[l_j(x) - l_j(0) - x l_j'(0) \right] \frac{1}{x^2}
\\
&
\quad
+ l_j(0)
\int_0^\infty \rmd x \,
w(x) \left[l_i(x) - l_i(0) - x l_i'(0) \right] \frac{1}{x^2}
\\
& 
\quad
+ \left[l_i(0) l_j'(0) + l'_i(0) l_j(0) \right]
 \int_0^\infty \rmd x \, \exp(-x^2)
\\
& 
\quad
+ l_i(0) l_j(0) 
 \int_0^\infty \rmd x \,
\frac{x}{x^2 + \eta^2} \exp(-x^2).
\end{split}
\end{equation}
The first three terms of the last equation can be computed using the
Gauss-Maxwell quadrature. The next term can be evaluated
analytically. The last integral can be computed analytically for small
$\eta$. We obtain
\begin{multline}
(B)_{ij}
= \frac{1}{x_i^2} \delta_{ij}
-  \left[\frac{1}{2} \gamma_E + \log(\eta) + 
\sum_{n=1}^{N} \frac{w_n}{x_n^2}
\right.
\\
\left.
+ (D_i + D_j) \left( \frac{1}{2} \sqrt{\pi} 
- \sum_{n=1}^{N} \frac{w_n}{x_n}\right) 
\right] l_i(0) l_j(0),
\label{eq:BijFinal}
\end{multline}
where $\gamma_E$ denotes the Euler-Mascheroni constant.

The last equation shows that $(B)_{ij}$ has a logarithmic singularity
for $\eta = 0$. For that reason we had to regularize the centrifugal
potential in Eq.~(\ref{eq:HspB}). An alternative way of treating such
singularity in the DVR method is discussed in Ref.~\cite{Vincke93}.

\end{document}